\RequirePackage{fix-cm}
\documentclass[smallextended]{svjour3}       
\smartqed  
\usepackage{lscape}
\usepackage[utf8]{inputenc}
\usepackage{graphicx}
\usepackage{caption}
\usepackage{subcaption}
\usepackage{fancyvrb}
\usepackage{float}
\usepackage{listings}
\usepackage{booktabs}
\usepackage{tabularx}
\usepackage{enumitem}
\usepackage{flushend}
\usepackage{url}
\usepackage[framemethod=tikz]{mdframed}
\usepackage{needspace}
\usepackage{xcolor}
\usepackage{mathptm}
\usepackage{multirow}

\mdfdefinestyle{mpdframe}{
    frametitlebackgroundcolor   =black!15,
    frametitlerule              =true,
    roundcorner                 =5pt,
    middlelinewidth             =1pt,
    innermargin                 =0.1cm,
    outermargin                 =0.1cm,
    innerleftmargin             =0.1cm,
    innerrightmargin            =0.1cm,
    innertopmargin              =0.1cm,
    innerbottommargin           =0.1cm
}

\newcommand{\RQone}{RQ1: How many changes and how much time do issues require to be addressed when SQL code is involved?}
\newcommand{\RQtwo}{RQ2: How many changes and how much time do pull requests require to be merged when SQL code is involved?}
\newcommand{\RQthree}{RQ3: Are the dimensions of effort invested in SQL development tasks different from those of non-SQL development tasks?}

\def\BibTeX{{\rm B\kern-.05em{\sc i\kern-.025em b}\kern-.08em
    T\kern-.1667em\lower.7ex\hbox{E}\kern-.125emX}}

\begin{document}

\title{Studying the Characteristics of SQL-related Development Tasks
}
\subtitle{An Empirical Study}


\author{Daniel Alencar da Costa         \and
        Natalie Grattan \and
	Nigel Stanger \and 
	Sherlock A. Licorish 
}


\institute{Daniel Alencar da Costa \at
              University of Otago \\
              Department of Information Science\\
              \email{danielcalencar@otago.ac.nz}           
	      \and
	      Natalie Grattan \at
              University of Otago \\
	      Department of Information Science\\
	      \email{natalie.grattan@postgrad.otago.ac.nz}
	      \and
	      Nigel Stanger \at
	      University of Otago\\
	      Department of Information Science\\
	      \email{nigel.stanger@otago.ac.nz}
	      \and
	      Sherlock A. Licorish \at
	      University of Otago\\
	      Department of Information Science\\
	      \email{sherlock.licorish@otago.ac.nz}
}

\date{Received: date / Accepted: date}

\maketitle


\begin{abstract}
	 A key function of a software system is its ability to facilitate the manipulation of data, which is often implemented using a flavour of the Structured Query Language (SQL).
	To develop the data operations of software (i.e, creating, retrieving, updating, and deleting data), developers are required to excel in writing and combining both SQL and application code. The problem is that writing SQL code in itself is already challenging (e.g., SQL anti-patterns are commonplace) and combining SQL with application code (i.e., for SQL development tasks) is even more demanding.
	Meanwhile, we have little empirical understanding regarding the characteristics of SQL development tasks. Do SQL development tasks typically need more code changes? Do they typically have a longer time-to-completion? Answers to such questions would prepare the community for the potential challenges associated with such tasks. 
	Our results obtained from 20 Apache projects reveal that SQL development tasks have a significantly longer time-to-completion than SQL-unrelated tasks and require significantly more code changes. Through our qualitative analyses, we observe that SQL development tasks require more spread out changes, effort in reviews and documentation. Our results also corroborate previous research highlighting the prevalence of SQL anti-patterns. The software engineering community should make provision for the peculiarities of SQL coding, in the delivery of safe and secure interactive software.

	\keywords{SQL anti-patterns \and SQL code \and SQL development task \and Software Maintenance}

\end{abstract}

\section*{Data Availability Statement}

The data that support our study's findings are openly available on Zenodo at \url{https://zenodo.org/record/7343828#.Y55ttXZBxPZ}

\section{Introduction}\label{sec:introduction}

Database storage and operations are key components of any software project requiring data persistence, ranging from small to large scale applications. Relational database management systems (RDBMSs) normally use the {\em Structured Query Language} (SQL) to access and manipulate data~\cite{Melton:1993}. SQL has reached such a degree of importance that despite developments in NoSQL DBMSs, SQL still remains a popular data manipulation mechanism~\cite{Gaspar:2017}. Evidence of the importance of SQL is that some NoSQL DBMSs have adapted to support SQL or SQL-like languages \cite{Gaspar:2017}. Other researchers have explored ways of bridging NoSQL to SQL through middleware~\cite{Rith:2014,Mason:2005}. Altogether, SQL code consistently remains a vital part of persistent data storage and is used for a variety of tasks. Thus, poorly written SQL can be the root cause of performance issues in applications \cite{Faroult:2006}, especially in resource-constrained environments, such as in mobile apps~\cite{Lyu:2019}. 

Recognizing the importance of SQL code, researchers have studied SQL in general~\cite{Edmundson:2013,Muse:2020,Karwin:2010,Almeida:2019}. A main line of research is the study of {\em SQL anti-patterns}, which are
recurrent mistakes when writing SQL~\cite{Nagy:2015}. The term SQL anti-pattern can also be interchangeably used with {\em SQL bad smell} or {\em SQL code smell}~\cite{Muse:2020}. 

The prevalence of SQL anti-patterns~\cite{Muse:2020,Almeida:2019} hints that developing SQL is a complex task within the already cognitive-heavy task of software development~\cite{Walenstein:2002,Mens:2012}. Further evidence that writing SQL code is not trivial is that there are roles dedicated to the development of SQL code (e.g., database administrators)~\cite{Yilmaz:2015,Miller:2008}. On the other hand, a full-stack developer, who writes and mixes both SQL and application code, faces the unique challenge of excelling in both worlds when performing {\em SQL development tasks} (i.e., development tasks that involve both application and SQL code). However, we have little empirical knowledge regarding the characteristics of SQL development tasks. For example, do SQL development tasks require more code changes than non-SQL development tasks? Do SQL development tasks take longer to be completed than non-SQL development tasks? Do SQL development tasks require different dimensions of development effort? Knowing these characteristics is important to inform software development tools and practices. For instance, if SQL development tasks typically require more code changes, {\em quality assurance} (QA) teams may use this information to adjust their code-reviewing priorities. As observed by Kononenko et al.~\cite{Kononenko:2018}, code size has a statistically significant association with code review time. Another area that could potentially benefit from our study is the area of software estimation~\cite{Usman:2014}, as understanding the characteristics of SQL development tasks could benefit the reasoning behind estimating story points for user stories that involve SQL. Finally, empirical research like ours is important to better understand special characteristics of atypical software development tasks. In this regard,  our study strives to better understand software development tasks, but focuses on SQL development tasks given their importance.

To investigate the characteristics of SQL development tasks, we performed an exploratory {\em Mining Software Repository} (MSR) study informed by qualitative document analyses~\cite{Bowen:2009,Oleary:2017}. Through a study of 20 carefully selected Apache projects, we performed comparisons between SQL and non-SQL development tasks. In the quantitative part of the study, we investigated two characteristics of SQL development tasks: {\em time-of-completion} and {\em size of changes}~\cite{Mende:2010,Kamei:2010,Weiss:2007,Giger:2010}. In the qualitative part of our study, we analyzed a stratified sample of 687 issue reports to better understand the nature of efforts invested in both SQL and non-SQL development tasks. 

\begin{itemize}[topsep=0pt,itemsep=0pt,leftmargin=*]
	\item \textbf{\RQone} 
Issue reports can describe bugs, enhancements, or new features to be addressed~\cite{daCosta:2014}. Among our main observations are the fact that SQL-related issues may take slightly longer to be addressed when compared to SQL-unrelated issues. Regarding the size of changes, we observe a clear trend that SQL-related issues require significantly more changes than SQL-unrelated issues. 
		\vspace{1mm}
	\item \textbf{\RQtwo}  
Another core activity in software development, especially in open source projects, is the review of pull requests (also known as merge requests). Our results do not reveal a clear trend whether SQL-related pull requests have a longer time-of-completion when compared to SQL-unrelated pull requests. Although most projects reveal that SQL-related pull requests take longer to be completed, two projects show the opposite trend. Regarding the change size of SQL-related and SQL-unrelated pull requests, we did not observe significant differences in our data. 
		\vspace{1mm}
	\item \textbf{\RQthree} 
		We qualitatively investigate a representative sample of 687 issue reports to better understand the dimensions of effort invested in SQL-related and SQL-unrelated tasks. Our main observation is that SQL-related task are more likely to have a larger scope when compared to SQL-unrelated tasks.
\end{itemize}

Our paper is organized as follows. In Section~\ref{sec:relatedwork}, we survey the research related to our work. In Section~\ref{sec:methodology}, we describe the methodology of our empirical study. We present our obtained results in Section~\ref{sec:results} and discuss them in Section~\ref{sec:discussion}. In Section~\ref{sec:threats}, we reflect on the threats to validity of our study and we conclude our paper in Section~\ref{sec:conclusion}.

\section{Related Work}\label{sec:relatedwork}

In this section, we explore two research themes that are closely related to our work: SQL anti-patterns spanning tools or anti-pattern automators, and empirical studies analysing SQL code and anti-patterns.

\subsection{Tools or Anti-Pattern Automators}
Identifying anti-patterns in source code is a time consuming task. Therefore, much research have invested in the creation of new tools to help automate the process of identifying SQL or {\em Object-Relational Mapping} (ORM) anti-patterns.

Chen et al.~\cite{Chen:2014} proposed a framework for detecting anti-patterns in ORM usage. They also identified the most common anti-patterns that have a performance impact. These anti-patterns are ``one-by-one processing'' and ``excessive data''~\cite{Chen:2014}. In a follow-up study, Chen et al.~\cite{Chen2:2016} investigated the impact of redundant data on performance in Java applications, further emphasising how inefficiencies in data access, retrieval, and storage can impact performance in a system. Cheung et~al.~\cite{Cheung:2013} aimed to optimize how SQL is used in software projects. They proposed a tool that parses source code and locates areas where code could be moved into the database layer rather than being carried out in the application code, since it is a costly operation to communicate from the application to the database.  

Nagy et al.~\cite{Nagy2:2015} designed a tool for locating the parts of source code where SQL queries are sent to the database. The tool can be a potential solution to the problem of not being aware of the query that is executed by the database, which can mask the complexity of the query. Lyu et al.~\cite{Lyu:2018} investigated the {\em Repetitive Autocommit Transaction} (RAT) anti-pattern, which involves repeated database transactions rather than batch transactions in mobile applications. This anti-pattern can have a significant impact on performance, so they proposed a tool which identifies this anti-pattern and refactors instances that can be changed without causing deadlocks.  

Much research in this area involve SQL static code analyzers, which automate the extraction of SQL code or anti-patterns from a source code repository. The tool Alvor is an Eclipse IDE plug-in that checks whether SQL statements embedded into code are syntactically correct, through a static analysis of the code to find SQL statements~\cite{Annamaa:2010}. Wasserman~et~al.~\cite{Wassermann:2007} presented a technique to evaluate dynamic SQL queries.
\texttt{SQLInspect}, designed by Nagy et al.~\cite{Nagy:2018}, is another Eclipse plug-in that extracts SQL queries and finds anti-patterns (based on Karwin's~\cite{Karwin:2010} anti-patterns). The authors used benchmark projects and compared the performance against Alvor and other similar tools. \texttt{SQLCheck}, developed by Dintyala et al.~\cite{Dintyala:2020}, is a similar tool that extracts SQL queries and detects anti-patterns. However, \texttt{SQLCheck} adds further functionality by ranking the anti-patterns based on their impact in the application, and fixing the anti-patterns by suggesting alternative approaches. The authors used SQL queries from real-world contexts to evaluate \texttt{SQLCheck}. 

In contrast to previous research, we are not interested in detecting anti-patterns in SQL code, but in understanding whether development tasks that involve both SQL and application code (i.e., SQL development tasks) have special characteristics (e.g., a longer time-to-completion or different dimensions of effort). These investigations can help us further improve software development processes and practices.

\subsection{Empirical Studies on SQL Anti-Patterns}
Recent studies have conducted empirical analyses of SQL anti-patterns.  
Lyu et al.~\cite{Lyu:2019} examined the impact of SQL anti-patterns on code performance in mobile applications. To compile a list of anti-patterns, the authors conducted an extensive literature review. They then used a benchmark design with values for performance-affecting factors and anti-pattern instances from a sample of Google Play applications. They tested whether performance improved by measuring changes in resource consumption. They found the anti-patterns ``unbatched-writes'' and ``loop-to-join''~\cite{Lyu:2019} to be prominent in terms of performance impact.  

The anti-patterns outlined by Karwin~\cite{Karwin:2010} provided a turning point in SQL anti-pattern research. Eessaar et al.~\cite{Eessaar:2015} investigated 12 of Karwin's SQL anti-patterns, both logical and physical. To investigate the anti-patterns, they ran SQL queries based on a test database with design flaws and investigated the {\em Information Schema} tables to analyze the design of the database. The authors were able to identify the majority of the anti-patterns. The purpose of the study was not to test the performance impact of these anti-patterns but rather find methods for detecting them.

Nagy et al.~\cite{Nagy:2015} parsed a dataset of \textsc{Stack Overflow} posts related to MySQL. They extracted code snippets and filtered for SQL keywords. Next, they used a pattern detector, which involved breaking the elements into a tree structure and sorting nodes into buckets based on their similarity. The authors did not find particular patterns that were likely to lead to errors. 

Another source of data for empirical investigations is OSS projects hosted on \textsc{GitHub}.\footnote{https://github.com/} Studies conducted by Almeida et al. \cite{Almeida:2019} and Muse et al. \cite{Muse:2020} used \textsc{GitHub} to collect their data. Almeida et al.~\cite{Almeida:2019} investigated SQL anti-patterns in PL/SQL projects. They studied 20 popular \textsc{GitHub} projects and found that certain anti-patterns tended to be highly correlated with one another and that they tended to fall into clusters around syntax bad smells and code structure bad smells. 
Muse et al.~\cite{Muse:2020} used \texttt{SQLInspect} \cite{Nagy:2018} to extract SQL anti-patterns. They also extracted traditional code smells to measure correlations between SQL anti-patterns and traditional code smells. They found that ``implicit columns'' was the most common SQL anti-pattern. They also found that SQL anti-patterns tended to persist throughout iterations of the project. Given that Muse et al.~\cite{Muse:2020} used an automated approach to identify SQL anti-patterns, their investigations were limited to only four anti-patterns. 

The existing empirical research on SQL anti-patterns has shown that SQL anti-patterns are indeed prevalent. However, what is missing is a closer analysis regarding the development process when it comes to SQL development tasks. For instance, are developers exerting as much effort when it comes to performing SQL development tasks? Our study, for the first time, provides an in-depth analysis regarding the characteristics of SQL development tasks. Our qualitative analysis helps us better understand the effort invested in development activities involving SQL code.  

\section{Methodology}\label{sec:methodology}

In this section, we review our subject projects and explain how the data used in this study were collected. We subsequently outline our research questions and associated measures.

\subsection{Subject Projects}
To perform our empirical investigations, we studied Apache projects \cite{Apache:2019}. We selected Java Apache projects because Java has been the main language used by Apache. Apache projects have been frequently used in prior software engineering research \cite{Rigby:2007,Roberts:2006,Neto:2018,Vandehei:2021}. We chose Apache projects for our study as a strategy to avoid the inclusion of toy, personal, or non-engineered projects that can be found among \textsc{GitHub} projects \cite{Kalliamvakou:2014,Pickerill:2020}. 

Another goal related to studying Apache projects is to control for quality, since the Apache Foundation has a thorough process to select its contributors (i.e., committers).\footnote{\url{https://www.apache.org/foundation/getinvolved.html#become-a-committer}} For example, Apache works as ``{\em a meritocracy. Once someone has shown sufficient sustained commitment to a project by helping out and contributing work to it (and the ASF), the project may vote to invite that person to become a committer.}''\footnote{\url{https://infra.apache.org/new-committers-guide.html}} 
Moreover, Apache has led the development of open source software  for 22 years, including the implementation of {\em Java Specification Requests} (JSRs), such as the Java Persistence API, or simply JPA.\footnote{\url{https://www.jcp.org/en/jsr/detail?id=317}}\textsuperscript{,}\footnote{\url{http://openjpa.apache.org/}}

Provided that we can ensure a certain level of professionalism in software development by selecting Apache projects, the second aspect of our project selection is to study projects with a substantial number of SQL queries. In this regard, we aim to carefully select projects with a substantial number of meaningful SQL statements. We explain the steps to select such projects in the next subsection.

\subsection{Procedure and Data Collection}
\begin{figure*}[htb]
	\centering
	\includegraphics[width=\textwidth,keepaspectratio]{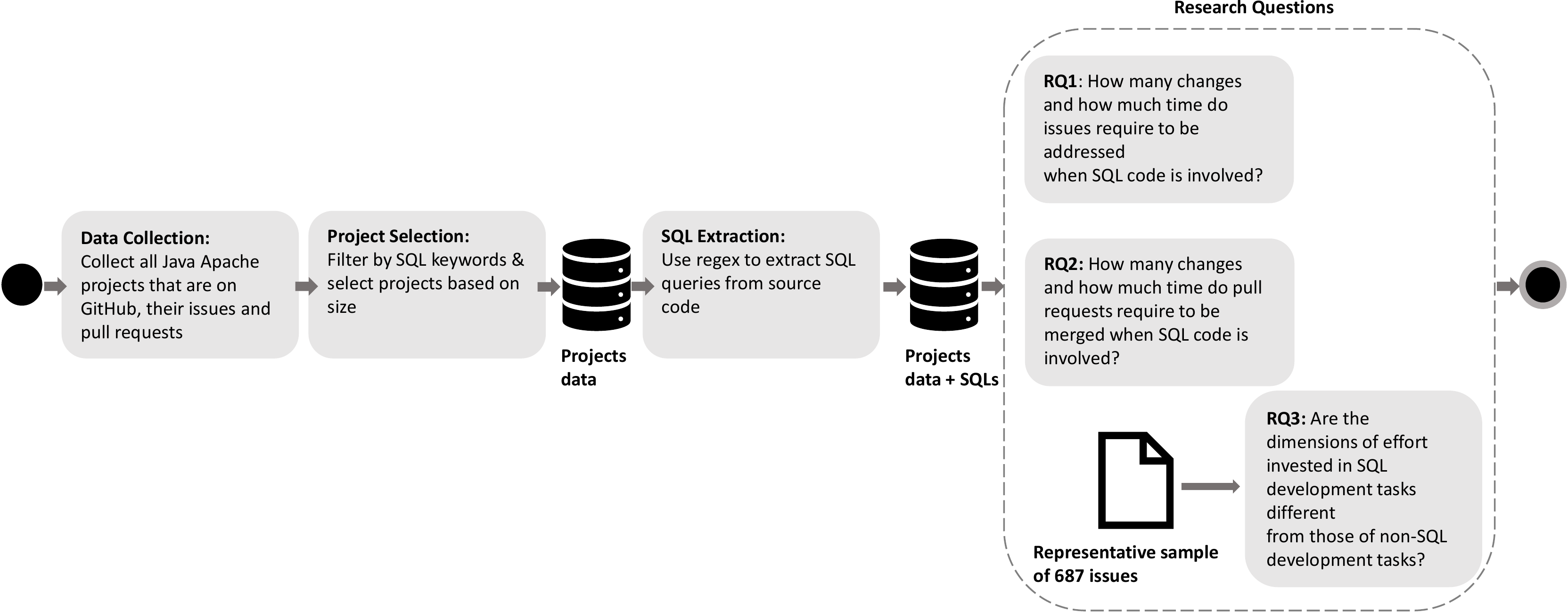}
	\caption{High-level overview of steps involved in our study.}
	\label{fig:researchemethod}
\end{figure*}

The first step to answer our research questions was to collect the required data. Each of the following steps refers to the corresponding heading in Figure \ref{fig:researchemethod}.

{\bfseries Step 1: Data Collection.} We used the \textsc{GHTorrent} database \cite{Gousi13} to obtain the names of all the Apache projects that have repositories available on \textsc{Github}. From here, we cloned each repository to a local computer and scanned each repository for SQL query keywords
including \texttt{SELECT}, \texttt{DELETE}, \texttt{UPDATE} and \texttt{INSERT}. This initial keyword scan of projects provided a baseline filter that we built upon later with additional regular expression filters. This formed a subset of Java Apache projects that are likely to contain SQL. The
commands to perform this initial scan can be found in our replication package.\footnote{\url{https://zenodo.org/record/7343828#.Y55ttXZBxPZ}} 

{\bfseries Step 2: Project Selection.} Similar to Almeida et. al. \cite{Almeida:2019}, we selected 20 Apache projects to perform our empirical study. We believe 20 is a reasonable number because our goal was to include all projects in our analyses, including the qualitative analyses of SQL development tasks and SQL anti-patterns in RQ2 and RQ3. 

Our protocol for sampling our 20 Apache projects is as follows. First, we found 2,100 projects related to Apache on \textsc{GitHub}. We excluded 1,100 projects as they were not mainly written in Java. We also excluded forks and kept only original repositories. Next, we scanned projects based on SQL keywords. Projects for which SQL keywords could not be found were excluded (778 projects were excluded). To select our final set of projects we counted the number of Java files in each of the remaining 208 projects. We observed that some projects contained an exorbitant number of files (e.g., Netbeans with about 37,500 files). As in this first research we do not intend to base our empirical observations on extreme cases, we removed outlier projects based on the upper limit of the inter quantile range of file counts, i.e., the same process used by boxplots to infer outliers (4 projects were removed). Finally, we selected the top 10 projects (in terms of file counts) and the 10 projects right above the median file count. This last step was intended to enrich our dataset with projects of different sizes and applications.



{\bfseries Step 3: SQL Extraction.} We developed a Python script to obtain all the string literals in the Java files of the selected projects. Our intention was to find queries written in native SQL. The script created a dataset of string literals, alongside details of what file and line number the string literal was from, and which project. The intention was to ensure that we could later revisit the context in which the SQL query was found, which is important for our analyses in answering RQ3.

In order to extract the SQL from the string literals, we crafted regular expressions. 
A different regular expression was made for each of: a \texttt{SELECT} statement, an \texttt{UPDATE} statement, an \texttt{INSERT} statement, and a \texttt{DELETE} statement. These were executed using the SQL regex tool to match the patterns.
The regular expressions were validated by the first and second authors using the \textsc{OODT} project, which is close to the median with 143 extracted SQL queries. A manual analysis was performed in this project by analyzing all the Java files and counting the number of SQL queries. Next, we compared the set of SQL queries from the manual analyses with those found by the regular expressions. The regular expressions performed well with a precision of 96.5\% and recall of 97.9\%. While we validated our regular expressions on only one Apache project, due to the high accuracy obtained, we are fairly confident that the majority of SQL queries can be  extracted. Moreover, we also performed another manual validation of the SQL queries found by our regular expressions at a later stage in answering RQ3.

Table~\ref{tbl:projectsstats} shows the statistics related to our selected projects, i.e., the number of SQL statements, number of SQL-related issues and pull requests (PRs), and number of SQL-unrelated issues and pull requests. Overall, our selected projects contain a mean of 722 SQL queries, with
a maximum of 6,010 queries (Hive) and a minimum of 5 queries (Tuscany). These numbers suggest that our studied projects are data-intensive systems \cite{Muse:2020}. 

\begin{landscape}
    \begin{table}
        \caption{General characteristics of our selected Apache projects. \label{tbl:projectsstats}}
    \end{table}
    \begin{center}
        \begin{tabular}{lrrrrr}
            \toprule
            \textbf{Project} & \textbf{SQL-Statements} & \textbf{SQL-rel. issues} & \textbf{SQL-unrel. issues} &
            \textbf{SQL-rel. PRs} & \textbf{SQL-unrel. PRs} \\
            \midrule
            \textsc{Cloudstack} & 1,049 & 93 & 2,329 & 353 & 7,599 \\ 
            \textsc{AsterixDB} & 18 & 4 & 187 & 2 & 16 \\
            \textsc{Geode} & 4,356 & 172 & 3,404 & 16 & 6 \\ 
            \textsc{Hadoop Map-Reduce} & 37 & 6 & 621 & -- & -- \\ 
            \textsc{Groovy} & 58 & 19 & 3,812 & 12  & 3,260 \\ 
            \textsc{Harmony} & 406 & -- & 110 & -- & -- \\ 
            \textsc{Hive} & 6,010 & 697 & 6,803 & 556 & 1,840 \\ 
            \textsc{Karaf} & 67 & 3 & 436 & 27 & 3,551 \\ 
            \textsc{Tomee} & 230 & 5 & 263 & 57 & 1,955 \\ 
            \textsc{Flink} & 354 & 198 & 5,254 & 269 & 5,111 \\ 
            \textsc{Knox} & 11 & 5 & 1,124 & 16 & 1,216 \\ 
            \textsc{Lucene-Solr} & 325 & 116 & 9,301 & 49 & 3655 \\ 
            \textsc{ManifoldCF} & 108 & --  & 4 & -- & 23 \\ 
            \textsc{Marmotta} & 34 & 2 & 253 & 2 & 6 \\ 
            \textsc{Nifi} & 933 & -- & 31 & 193 & 3,726\\ 
            \textsc{OODT} & 143 & 2 & 24 & 5 & 121 \\ 
            \textsc{Synapse} & 24 & -- & 42 & 1 & 32 \\ 
            \textsc{Oozie} & 262 & 97 & 1,364 & 12 & 28 \\ 
            \textsc{OpenWebBeans} & 6 & 2 & 724 & -- & 5 \\ 
            \textsc{Tuscany} & 5 & 4 & 835 & --  & -- \\ 
            \bottomrule
        \end{tabular}
        \footnote{Note that some projects, such as Harmony and Tuscany, did not have any pull requests in their \textsc{GitHub} repositories at the time our dataset was collected.}

    \end{center}
\end{landscape}

{\bfseries Step 4: Data Mining.} After obtaining our list of projects, we collected additional data related to pull requests (from \textsc{GitHub}) and issue reports (from \textsc{JIRA}). We collected pull requests marked as {\em closed} or {\em closed and merged} and considered only pull requests that have been merged to the main branch for the quantitative analyses as we are interested in computing time-of-completion and change size metrics only for those pull requests that eventually made it to the end users. Additionally, even if any of those pull requests involved other branches, because the work has eventually been merged to the main branch, we did not lose information when computing the {\em time-to-completion} for such pull requests. Regarding issue reports, we collected reports marked as closed and fixed. This is because our goal is to investigate the time taken for pull requests and issue reports to be marked as closed. These collection tasks were facilitated using Python libraries such as \texttt{PyGithub} and \texttt{jira}. 

\subsection{Research Questions}\label{subsec:rqs}
We reiterate below our three research questions along with their motivations, approaches and measures.

\textit{\bfseries \RQone}
Issue reports can describe bugs, enhancements, or new features to be addressed~\cite{daCosta:2014}. Therefore, addressing issue reports sits at the core of the software development endeavour. We study the time-of-completion and size of issue reports to better understand whether there are significant differences when issue reports that involve SQL code. This is our preliminary investigation to understand whether SQL development tasks are addressed in a different manner when compared to non-SQL development tasks. 

\textit{\bfseries \RQtwo}
Another core activity in software development, especially in open source projects, is the review of pull requests (also known as merge requests). Pull requests are normally reviewed by other (typically more experienced) members of the project development team \cite{Rahman:2014}. In RQ2, we study time-of-completion and size of pull requests to better understand whether there are significant differences when pull requests involve SQL code. As in RQ1, RQ2 help us better understand whether SQL development tasks are addressed differently when compared to non-SQL development tasks.

\textit{\bfseries \RQthree}
We further investigated our data to better understand why the {\em time-of-completion} and {\em change size} of SQL development tasks significantly differ from those of non-SQL development tasks. This investigation is important not only to triangulate the results obtained in RQ1 but also to provide deeper insights for researchers and practitioners. For example, although the {\em time-to-completion} of SQL development tasks is typically longer when compared to non-SQL development tasks, the difference could arise either because (i) SQL development tasks are less important, taking longer to receive attention from developers; or because (ii) SQL development tasks require more effort from developers, thus taking longer to complete. The goal of RQ3 is to shed more light on discussions of a similar nature as (i) and (ii). 

\section{Results}\label{sec:results}

\subsection*{\bfseries\RQone}

\noindent{\bfseries Approach.} To find which issues were related to SQL, we linked \textsc{GitHub} commits to issue reports from JIRA. We used an approach repeatedly used in previous research \cite{Posnett:2011,daCosta:2016,Neto:2018,Yatish:2019} to link JIRA issues with commits. For example, for the project OODT, we used Python scripts to search each commit message for ``OODT-'' followed by a number, which is the convention used by Apache projects to refer to issue IDs within commit logs.\footnote{\url{https://issues.apache.org/jira/browse/OODT-1031}}  Afterwards, for each commit, we ran our SQL regular expressions on only the patches for Java files. If any SQL queries were found within a patch, the issue was marked as containing SQL. To measure the {\em time-to-completion} of issue reports, we measured the difference between the date that an issue report was opened and the date that the issue report was fixed. To compute the {\em size} of an issue report, we summed all the added and removed lines in the commits linked to that issue report.

We used beanplots to visualize the distributions of {\em time-to-completion} and {\em size} \cite{Kampstra:2008}. To check whether distributions were statistically different (i.e., between SQL issues vs.\ non-SQL issues ), we used the Mann Whitney Wilcoxon (MWW) test \cite{Wilks:2011} and the Cliff's delta effect-size measurement \cite{Cliff:1993}. Both MWW and Cliff's delta are non-parametric, as Shapiro-Wilks tests indicated the non-normality of our data \cite{Shapiro:1965}. The MWW test checks whether two distributions come from the same population and the Cliff's delta measures the probability that a randomly selected value of one distribution is higher (or lower) than another distribution.\\

\noindent{\bfseries Results. \textit{SQL-related issues have a slightly longer time-to-completion when compared to SQL-unrelated issues.}}
Figure~\ref{fig:timetocompletionbeanplot} shows the distributions of time-to-completion for SQL-related issues versus SQL-unrelated issues. The left-hand distribution of each beanplot represents issues not involving SQL, whereas the right-hand distribution represents issues that involve SQL. We observe that, in terms of {\em time-to-completion}, SQL-related issues take slightly longer to be completed. For issues, we obtained $p = 5.33 \times 10^{-54}$ and Cliff's delta $d = 0.241$, indicating a small but significant difference. 

\begin{figure}
	\centering
	\includegraphics[width=.6\textwidth,keepaspectratio]
	{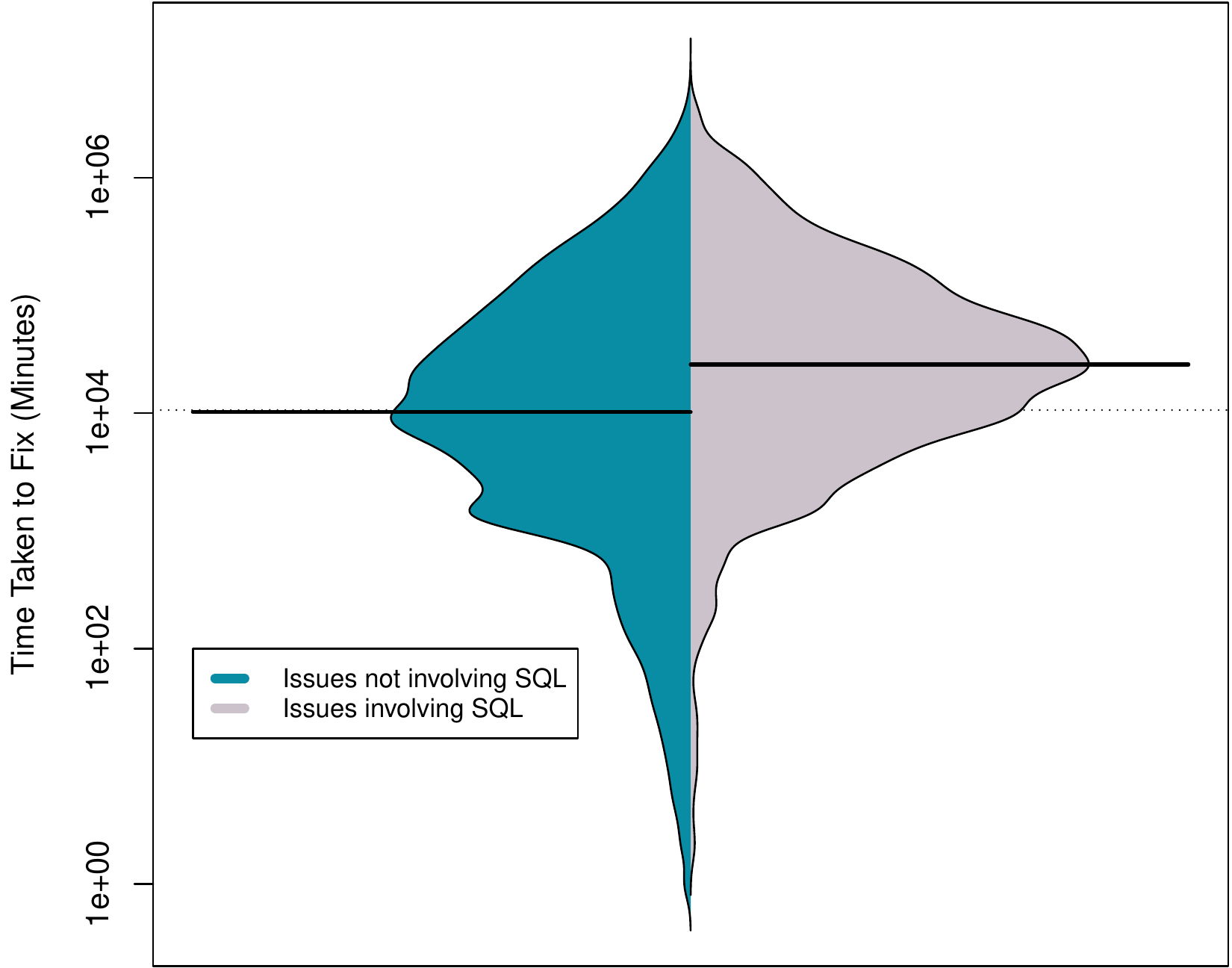}
	\caption{Time-to-completion in minutes. SQL-related issues and pull requests both take longer to be fixed in our studied projects. \label{fig:timetocompletion:issues}}
\end{figure}

\begin{table}[htb]
    \centering
	\begin{tabular}{lll}
		\toprule
            & \multicolumn{2}{c}{\textbf{ Time-to-Completion of Issues (All projects)}} \\
        \cmidrule(r){2-3}
            & \emph{SQL-related} & \emph{SQL-unrelated}  \\
		\addlinespace
		Minimum   & 2                    & 0                     \\
		Median    & 25,897 (17.9 days)   & 10,161 (7 days)       \\
		Mean      & 121,427 (84 days)    & 93,368 (65 days)      \\
		Maximum   & 3,888,797            & 6,129,740             \\
		\bottomrule
	\end{tabular}
	\caption{Descriptive statistics for time-to-completion (in minutes).\label{tbl:timetocompletion}}
\end{table}

\begin{figure}
	\centering
	\includegraphics[width=.6\textwidth,keepaspectratio]
	{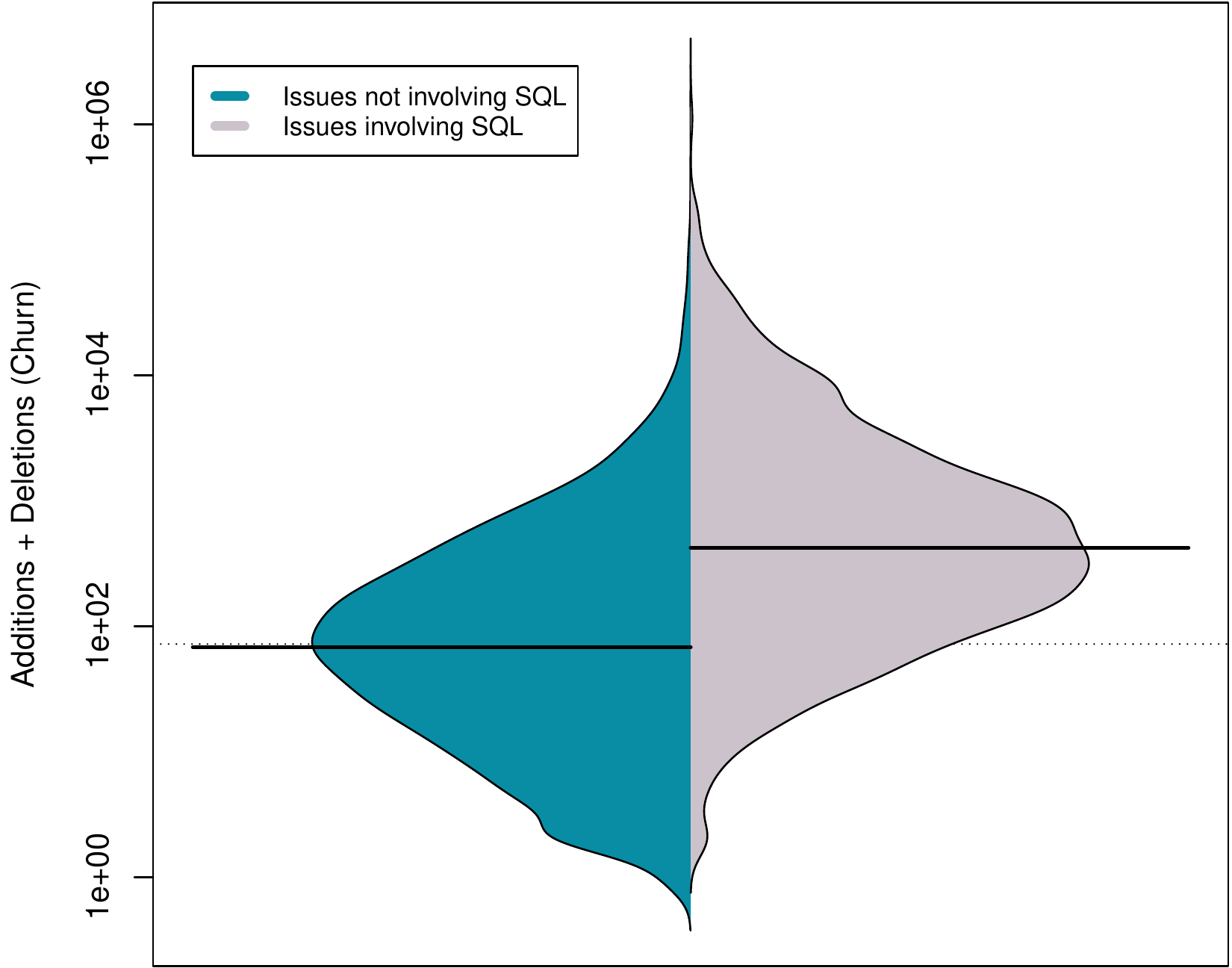}
	\caption{The size of changes (additions plus deletions). SQL-related issue reports receive more code changes. \label{fig:churn:issues}}
\end{figure}

\begin{table}[htb]
    \centering
	\begin{tabular}{lll}
		\toprule
            & \multicolumn{2}{c}{\textbf{Size of Issues (All projects)}} \\
        \cmidrule(r){2-3}
            & \emph{SQL-related} & \emph{SQL-unrelated} \\
		\addlinespace
		Minimum   & 2                    & 0         \\
		Median & 422		      & 68           \\
		Mean   & 4,338	      	      & 790          \\
		Maximum   & 1,115,413            & 1,810,320 \\

		\bottomrule
	\end{tabular}
	\caption{Descriptive statistics for change size (number of lines).\label{tbl:changesize}}
\end{table}

\noindent\textit{\bfseries SQL-related issues involve substantially more code modifications than other issues}
Figure~\ref{fig:churn:issues} shows the distributions of change size (i.e., the sum of code additions and code deletions) for SQL-related issues versus SQL-unrelated issues. We observe that SQL-related issues receive significantly more code modifications than SQL-unrelated issues: $p = 1.45 \times 10^{-230}$ and $d = 0.507$, indicating a {\em large} significant difference between the distributions. 

Table~\ref{tbl:changesize} shows descriptive statistics for the distributions shown in Figure~\ref{fig:churn:issues}. In line with the results above, we observe a large difference between the medians when it comes to issue reports: 422 lines modified for SQL-related issues against 68 lines modified for SQL-unrelated issues. These results suggest that SQL-related issues typically involve more code modifications than other issues.\\

\noindent{\bfseries Per project analysis (time-to-completion).} Our observations made so far concern the general trend of our data, i.e., we considered all projects in an aggregated manner. Although it is valuable to know the general trend in the data, it is also important (especially in software engineering) to understand the specificities and contextual profiles of each project, since each project uses different processes, tools and are from different domains. 

Table~\ref{tbl:timetocompletion:issues:perproject} shows the median time-to-completion of SQL-related issues versus SQL-unrelated issues in each of our studied projects. Table~\ref{tbl:timetocompletion:issues:perproject} also shows whether the results are statistically significant (i.e., as indicated by the MWW tests and Cliff's delta values). Four projects (\textsc{Nifi}, \textsc{Synapse}, \textsc{Harmony}, and \textsc{ManifoldCF}) were not included in the per-project analysis as we could not find SQL-related issues in these projects.

\begin{table}[htb]
    \centering
	\begin{tabular}{lllll}
		\toprule
	     \multicolumn{5}{c}{\textbf{Median time-to-completion of issues in days}}    \\
        \cmidrule(r){1-5}
	     & \emph{SQL-unrelated} & \emph{SQL-related} & \emph{p-value} & \emph{Cliff's delta} \\
		\addlinespace
		\textsc{AsterixDB} & 18.4  & 83.2 & 0.29 & --- \\
		\addlinespace
		\midrule
		\textsc{CloudStack} &  6 & 4.6 & 0.92 & ---  \\
		\addlinespace
		\midrule
		\textsc{Geode}  & 9.9 & 20.7 & \textbf{7.82$e^{-09}$} & \textbf{0.26 (small)}        \\
		\addlinespace
		\midrule
		\textsc{Groovy}  & 7.1 	& 19.9 & 0.07 & ---  \\
		\addlinespace
		\midrule
		\textsc{Hadoop-MapReduce}  & 27.9  & 51.3 & 0.20 & ---  \\
		\addlinespace
		\midrule
		\textsc{Hive}  & 7.1 & 17.9 & \textbf{9.01$e^{-32}$} & \textbf{0.27 (small)}        \\
		\addlinespace
		\midrule
		\textsc{Flink}  & 8.9 & 14.6 & \textbf{0.003} & 0.12 (negligible)        \\
		\addlinespace
		\midrule
		\textsc{Karaf}  & 2.9 & 0.4 & 0.86 & ---   \\
		\addlinespace
		\midrule
		\textsc{Knox}  & 0.1 & 39.5 & 0.09 & ---  \\
		\addlinespace
		\midrule
		\textsc{Lucene-Solr}  & 5.8  & 48.8 & \textbf{8.40$e^{-13}$} & \textbf{0.39 (medium)}        \\
		\addlinespace
		\midrule
		\textsc{Marmotta}  & 7 	& 1.5	& 0.41 & ---  \\
		\addlinespace
		\midrule
		\textsc{OODT}  &  2.2 & 52 & 0.89 & ---  \\
		\addlinespace
		\midrule
		\textsc{Oozie}  & 15  & 26.9 & \textbf{0.01} & \textbf{0.15 (small)}       \\
		\addlinespace
		\midrule
		\textsc{OpenWebbeans}  & 1.9 & 62.2 & 0.49 & ---        \\
		\addlinespace
		\midrule
		\textsc{Tomee}  & 0.3  	& 12.5	& 0.32 &  ---       \\
		\addlinespace
		\midrule
		\textsc{Tuscany} &  4 & 28.1 & 0.11 & ---      \\
		\addlinespace
		\bottomrule
	\end{tabular}
	\caption{Time-to-completion of issues per project.\label{tbl:timetocompletion:issues:perproject}}
\end{table}

Out of 16 analysed projects, 5 projects obtained significant p-values (\textsc{Geode}, \textsc{Hive}, \textsc{Flink}, \textsc{Lucene-Solr}, and \textsc{Oozie}). \textsc{Flink} obtained the smallest effect-size (i.e., considered negligible) whereas \textsc{Lucene-Solr} obtained the highest effect-size (i.e., considered medium). Although not all projects obtained a significant p-value, 12 out of 16 projects had a longer time-to-completion when it comes to SQL-related issues, which may explain why the general trend of time-to-completion (i.e., considering all the projects together) indicated a {\em small} but significant difference in the time-to-completion between SQL-related issues and SQL-unrelated issues. Overall, our per-project analysis suggests that the time-to-completion of SQL-issues may be slightly higher than SQL-unrelated issues in certain projects.\\

\noindent{\bfseries Per project analysis (size of changes).} In Figure~\ref{fig:churn:issues}, we observed a {\em large} significant difference in the size of changes between SQL-related issues and SQL-unrelated issues. To gain further insights, we study the differences in the size of changes in each of our analysed projects. Table~\ref{tbl:churn:issues:perproject} shows the median change size (for both SQL-related and SQL-unrelated issues) as well as the p-values obtained through our statistical tests.

\begin{table}[htb]
    \centering
	\begin{tabular}{lllll}
		\toprule
	     \multicolumn{5}{c}{\textbf{Median size of addressed issues}}    \\
        \cmidrule(r){1-5}
	     & \emph{SQL-unrelated} & \emph{SQL-related} & \emph{p-value} & \emph{Cliff's delta} \\
		\addlinespace
		\textsc{AsterixDB} & 208  & 3017  & \textbf{0.008} & \textbf{0.77 (large)} \\
		\addlinespace
		\midrule
		\textsc{CloudStack} & 18.4  & 83.2 & \textbf{3.23$e^{-13}$} & \textbf{0.45 (medium)} \\
		\addlinespace
		\midrule
		\textsc{Geode}  & 85 & 854 & \textbf{1.72$e^{-41}$} & \textbf{0.61 (large)}        \\
		\addlinespace
		\midrule
		\textsc{Groovy}  & 50 & 293 & \textbf{3.75$e^{-07}$} & \textbf{0.67 (large)} \\
		\addlinespace
		\midrule
		\textsc{Hadoop-MapReduce}  & 92  & 836 & 0.006  & \textbf{0.65 (large)}  \\
		\addlinespace
		\midrule
		\textsc{Hive}  & 77 & 406 & \textbf{4.61$e^{-84}$} & \textbf{0.45 (medium)}        \\
		\addlinespace
		\midrule
		\textsc{Flink}  & 82 & 403 & \textbf{1.37$e^{-28}$} & \textbf{0.46 (medium)}        \\
		\addlinespace
		\midrule
		\textsc{Karaf}  & 27 & 133 & 0.08 & ---   \\
		\addlinespace
		\midrule
		\textsc{Knox}  & 45 & 1,117 & 0.09 & ---  \\
		\addlinespace
		\midrule
		\textsc{Lucene-Solr}  & 90  & 529 & \textbf{0.001} & \textbf{0.80 (large)}        \\
		\addlinespace
		\midrule
		\textsc{Marmotta}  & 87 & 968 & 0.06 & ---  \\
		\addlinespace
		\midrule
		\textsc{OODT}  &  24 & 19,239 & \textbf{0.02}  & \textbf{0.90 (large)}  \\
		\addlinespace
		\midrule
		\textsc{Oozie}  & 41  & 384 & \textbf{1.36$e^{-24}$} & \textbf{0.62 (large)}       \\
		\addlinespace
		\midrule
		\textsc{OpenWebbeans}  & 91 & 16,241 & \textbf{0.01} & \textbf{0.97 (large)}        \\
		\addlinespace
		\midrule
		\textsc{Tomee}  & 92  	& 661	& \textbf{0.02} & \textbf{0.59 (large)}      \\
		\addlinespace
		\midrule
		\textsc{Tuscany} &  77 & 356 & 0.11 & ---  \\
		\addlinespace
		\bottomrule
	\end{tabular}
	\caption{Size of issues per project.\label{tbl:churn:issues:perproject}}
\end{table}

Our results revealed a clear trend of SQL-related issues requiring significantly more changes than SQL-unrelated issues. It is observed that 75\% of our projects ($\frac{12}{16}$) obtained a significant p-value with (at least) {\em medium} effect size measurements. Indeed, 9 projects obtained a {\em large} difference according to their effect size measurements. Our results suggest that SQL-related issues indeed require significantly more changes in order to be addressed.

\subsection*{\bfseries\RQtwo}

\noindent{\bfseries Approach.} To identify whether pull requests were related to SQL, we ran our regular expressions on the commits associated with each pull request (again, only on patches for Java files). To compute the {\em time-to-completion} of each pull request to be merged or closed, we computed the difference between the date that a pull request was submitted and the date it was closed or merged. To compute the {\em size} of a pull request, we summed all the added and removed lines in the commits associated with that pull request. As in RQ1, we used beanplots to visualize the distributions of {\em time-to-completion} and {\em size}. We also use the MWW test and the Cliff's delta measurement to compare our distributions.\\

\begin{figure*}
	\centering
		\includegraphics[width=.6\textwidth,keepaspectratio]
		{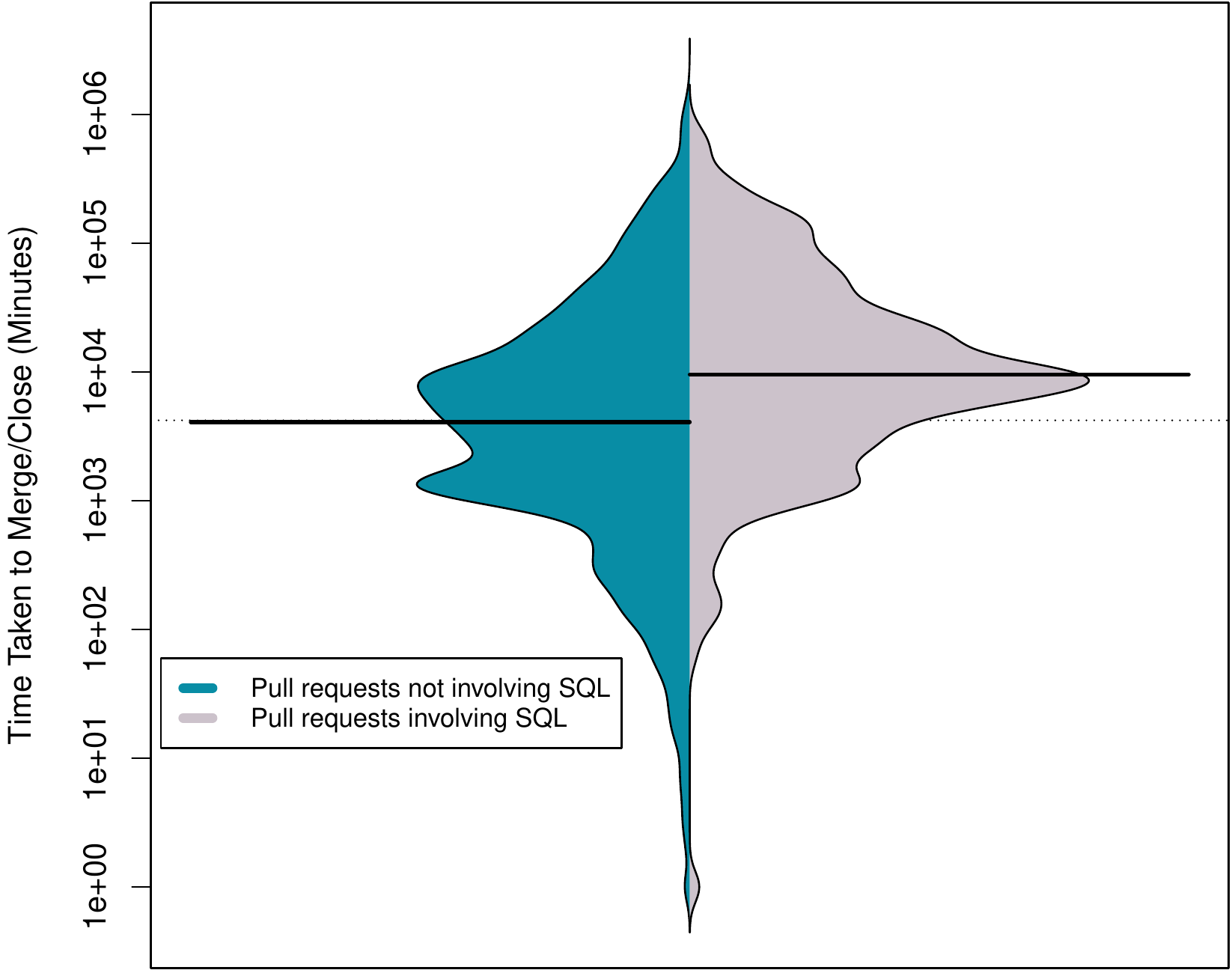}
		\caption{Pull requests}\label{fig:timetocompletion:pullrequests}
	\caption{Time-to-completion in minutes. SQL-related issues and pull requests both take longer to be fixed in our studied projects. \label{fig:timetocompletionbeanplot}}
\end{figure*}

\noindent{\bfseries Results. \textit{SQL-related pull requests have a slightly longer time-to-completion when compared to SQL-unrelated pull requests.}} 
Figure~\ref{fig:timetocompletion:pullrequests} compares the distribution of time-to-completion between SQL-related pull requests and SQL-unrelated pull requests.
We obtained a $p = 6.14 \times 10^{-22}$ and a $d = 0.28$, indicating a {\em small} but significant difference. The descriptive statistics for our distributions of time-to-completion are shown in Table~\ref{tbl:timetocompletion:pr}. Our results suggest that SQL-related pull requests may take slightly longer to be completed when compared to their SQL-unrelated counterparts.\\

\begin{table}[htb]
    \centering
	\begin{tabular}{lll}
		\toprule
            & \multicolumn{2}{c}{\textbf{ Time-to-Completion of Pull requests (All projects)}} \\
        \cmidrule(r){2-3}
            & \emph{SQL-related} & \emph{SQL-unrelated}  \\
		\addlinespace
		Minimum   & 0        		& 0                     \\
		Median    & 9,588 (7 days)      & 4,086 (3 days)       \\
		Mean      & 36,940 (26 days)	& 25,265 (17 days)      \\
		Maximum   & 757,374  		& 1,731,191             \\
		\bottomrule
	\end{tabular}
	\caption{Descriptive statistics for time-to-completion (in minutes).\label{tbl:timetocompletion:pr}}
\end{table}

\begin{figure}
	\centering
		\includegraphics[width=.6\textwidth,keepaspectratio]
		{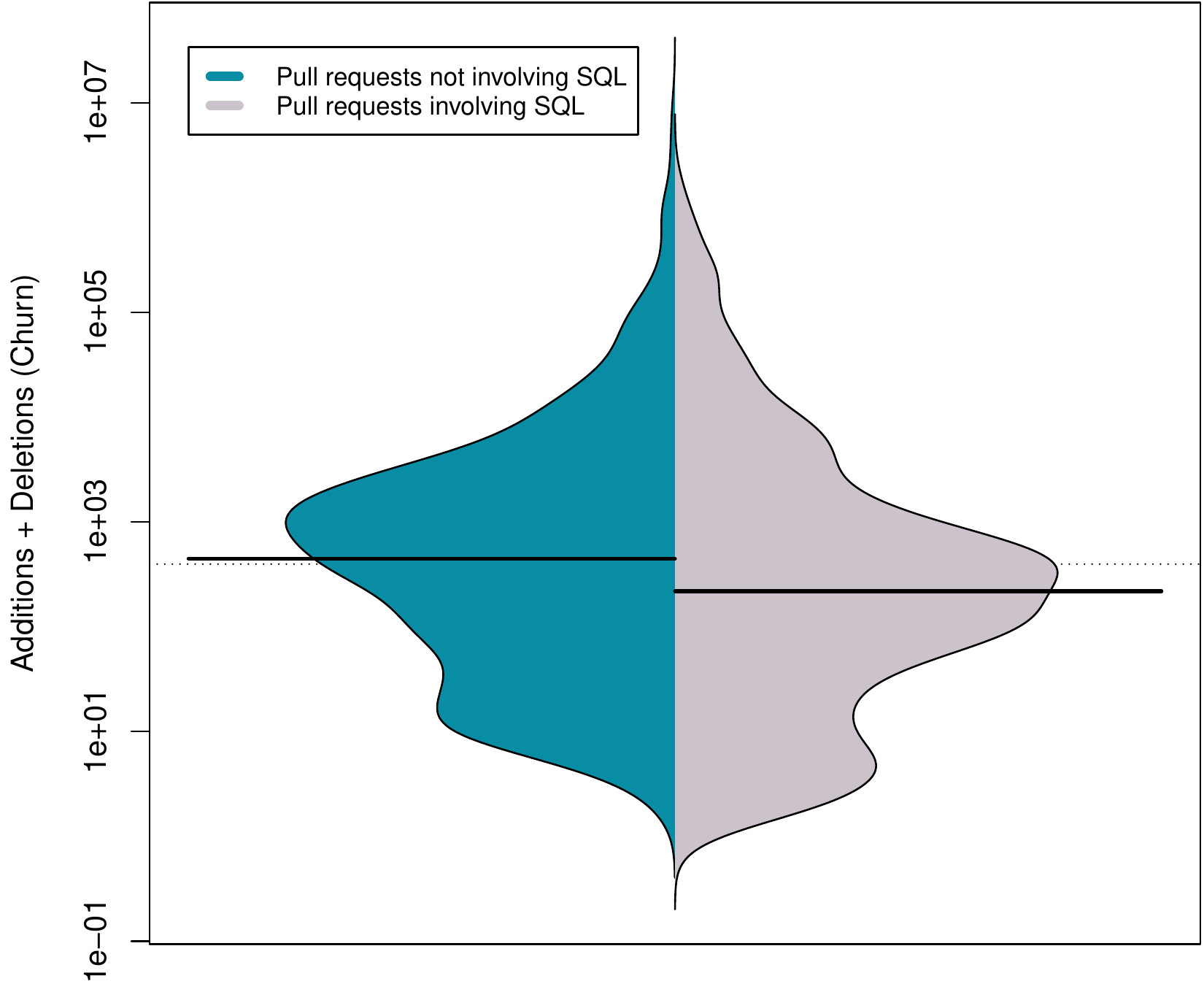}
		\caption{Pull requests}\label{fig:changesizebeanplot:pullrequests}
	\caption{The size of changes (additions plus deletions). SQL-related issue reports receive more code changes to be addressed. \label{fig:changesizebeanplot}}
\end{figure}

\noindent{\bfseries SQL-related pull requests have a negligible difference in terms of change size when compared to SQL-unrelated pull requests.}
Figure~\ref{fig:changesizebeanplot:pullrequests} shows the comparison of change size between SQL-related pull requests and SQL-unrelated pull requests (i.e., added lines plus removed lines within patches). We observe that SQL-unrelated pull requests received a slightly higher amount of code modifications compared to SQL-related pull requests. However, while the p-value for the comparison is significant ($p = 6.42 \times 10^{-5}$), our Cliff's delta measurement indicates a {\em negligible} difference ($d = 0.12$), meaning that the observed difference is likely inconclusive. Table~\ref{tbl:changesize:pr} shows the descriptive statistics of the size of changes for our pull requests.

\begin{table}[htb]
    \centering
	\begin{tabular}{lll}
		\toprule
            & \multicolumn{2}{c}{\textbf{Change size of Pull requests (All projects)}} \\
        \cmidrule(r){2-3}
            &  \emph{SQL-related} & \emph{SQL-unrelated} \\
		\addlinespace
		Minimum   &  0 & 1                        \\
		Median &  218 & 446          \\
		Mean   &  18,124 & 29,300         \\
		Maximum   &  1,563,226 & 8,434,616    \\
		\bottomrule
	\end{tabular}
	\caption{Descriptive statistics for change size (number of lines).\label{tbl:changesize:pr}}
\end{table}

\noindent{\bfseries Per project analysis (time-to-completion).} To better understand the general distribution of time-to-completion shown in Figure~\ref{fig:timetocompletion:pullrequests} we compare the distributions of time-to-completion in each studied project. In this per-project analysis, we did not include the \textsc{Tuscany}, \textsc{Hadoop-MapReduce}, \textsc{Harmony}, \textsc{ManifoldCF}, and \textsc{OpenWebbeans} projects. In the cases of \textsc{Tuscany}, \textsc{ManifoldCF}, and \textsc{OpenWebbeans}, we could not find SQL-related pull requests, whereas there were no closed pull requests in \textsc{Hadoop-MapReduce} and \textsc{Harmony}.
Table~\ref{tbl:timetocompletion:pr:perproject} shows our obtained results for the per-project analysis (i.e., median time-to-completion and p-values).

\begin{table}[htb]
    \centering
	\begin{tabular}{lllll}
		\toprule
	     \multicolumn{5}{c}{\textbf{Median time-to-completion of issues in hours}}    \\
        \cmidrule(r){1-5}
	     & \emph{SQL-unrelated} & \emph{SQL-related} & \emph{p-value} & \emph{Cliff's delta} \\
		\addlinespace
		\textsc{AsterixDB} & 4.4  & within 1 hour & 0.16 & --- \\
		\addlinespace
		\midrule
		\textsc{CloudStack} & 170  & 465 & \textbf{5.40$e^{-8}$} & \textbf{0.17 (small)}  \\
		\addlinespace
		\midrule
		\textsc{Geode}  & 1,179 & within 1 hour & \textbf{0.001} & \textbf{0.97 (large)}        \\
		\addlinespace
		\midrule
		\textsc{Groovy}  & 32 & 82 & 0.75 & ---  \\
		\addlinespace
		\midrule
		\textsc{Hive}  & 435 & 568 & 0.10 & ---  \\
		\addlinespace
		\midrule
		\textsc{Flink}  & 72 & 142 & \textbf{6.64$e^{-11}$} & \textbf{0.23 (small)}        \\
		\addlinespace
		\midrule
		\textsc{Karaf}  & 20 & 14 & 0.77 & ---   \\
		\addlinespace
		\midrule
		\textsc{Knox}  & 15 & 86 & \textbf{0.001} & \textbf{0.45 (medium)}   \\
		\addlinespace
		\midrule
		\textsc{Lucene-Solr}  & 121  & 2,801 & \textbf{1.09$e^{-13}$} & \textbf{0.37 (medium)}        \\
		\addlinespace
		\midrule
		\textsc{Marmotta}  & within 1 hour & within 1 hour & 1 & ---  \\
		\addlinespace
		\midrule
		\textsc{OODT}  &  53 & 125 & 0.30 & ---  \\
		\addlinespace
		\midrule
		\textsc{Oozie}  & 294  & 66 & \textbf{0.03} & \textbf{0.42 (medium)}       \\
		\addlinespace
		\midrule
		\textsc{Synapse}  & 30 & within 1 hour & \textbf{7$e^{-4}$} &  \textbf{0.98 (large)}      \\
		\addlinespace
		\midrule
		\textsc{Nifi}  & 52 & 186 & \textbf{2.53$e^{-16}$} &  \textbf{0.25 (small)}      \\
		\addlinespace
		\midrule
		\textsc{Tomee} &  48 & 175 & \textbf{0.01} & \textbf{0.18 (small)}      \\
		\addlinespace
		\bottomrule
	\end{tabular}
	\caption{Time-to-completion of pull requests per project.\label{tbl:timetocompletion:pr:perproject}}
\end{table}

We observed that 60\% of the analysed projects ($\frac{9}{15}$) obtained statistically significant difference. While the majority of these projects tend to have a longer time-to-completion for SQL-related pull requests, two projects (\textsc{Geode} and \textsc{Synapse}) show the opposite trend (i.e., SQL-unrelated pull requests have a longer time-to-completion than SQL-related pull requests). 

Overall, our results suggest that more projects need to be investigated to draw stronger conclusions related to the time-to-completion of pull requests. For example, reflecting on our previous analysis regarding issue reports, although not all projects obtained significant p-values, the results were consistent (i.e., all projects had a longer time-to-completion for SQL-related issue reports). However, when it comes to the per-project analysis of pull requests, \textsc{Geode} and \textsc{Synapse} revealed an opposite trend from the general trend shown in Figure~\ref{fig:timetocompletion:pullrequests}. Lastly, given that we did not observe a significant difference between SQL-related and SQL-unrelated pull requests when it comes to the size of changes, we do not perform a per-project analysis regarding the size of changes for pull requests. 

\subsection*{\bfseries\RQthree}

\noindent{\bfseries Approach.} To answer RQ2, we pushed beyond the quantitative realm and used a qualitative approach known as {\em document analysis} \cite{Bowen:2009,Oleary:2017}. Analyzing documents entails coding their content into themes (similar to how interview transcripts would be analyzed) \cite{Bowen:2009}. One of the main advantages of document analysis is that documents are ``non-reactive,'' meaning that documents can be read and revisited multiple times without being changed by the research process (\cite{Bowen:2009}, p.\ 31). As the input for our document analysis, we used {\em digital issue reports} and their related documents, e.g., pull requests (from all branches), code review boards, or commit logs related to the issue reports. From the total of 38,160 issue reports obtained from our 20 studied projects, we created two representative samples: one containing issue reports involving SQL code, and the other containing issue reports that do not involve SQL code. Considering a confidence level of 95\% and a confidence interval of 5\%, we obtained 304 issue reports involving SQL code and 383 issue reports not involving SQL code. Instead of simply building random samples, we used a {\em Stratified Random Sample} (SRS) strategy because of the variability in the number of issue reports involving SQL per project in our population~\cite{Lohr:2009}. Therefore, the number of issue reports involving SQL code per project in our samples is representative of the number of issue reports per project in the original population. 
To calculate the size of each stratum, we used the formula $n_{h} \approx n W_{h}$, where $n_{h}$ is the sample size of stratum $h$; $n$ is the size of the sample; and $W_{h} = N_{h}/N$ where $N$ is the population size and $N_h$ is the size of stratum $h$ in the population \cite{Podgurski:1999}. 

\begin{figure}
	\centering
		\includegraphics[width=.6\textwidth,keepaspectratio]
		{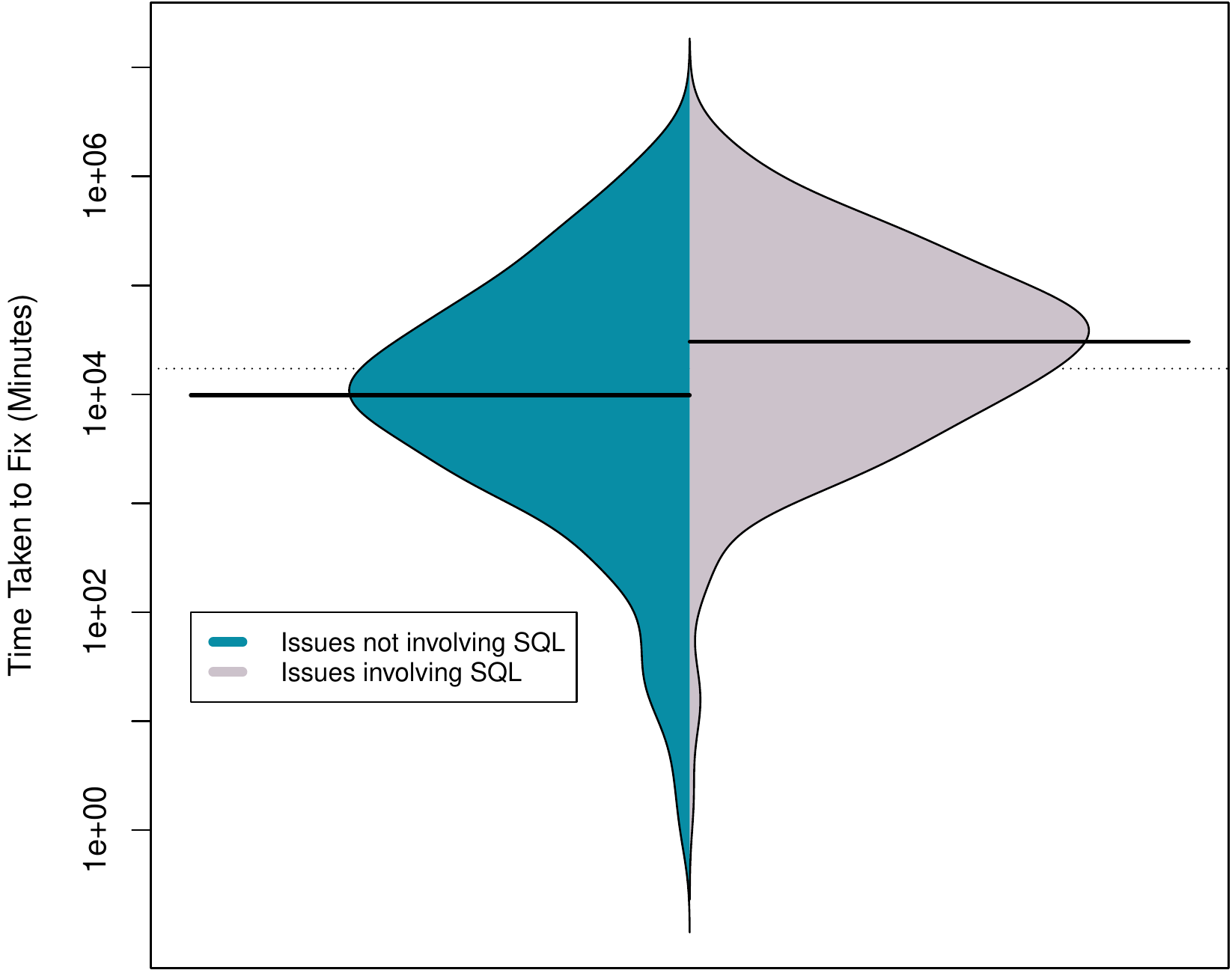}
	\caption{Time-to-completion of the issues in our representative sample. \label{fig:sample:timetocompletion}}
\end{figure}

\begin{figure}
	\centering
		\includegraphics[width=.6\textwidth,keepaspectratio]
		{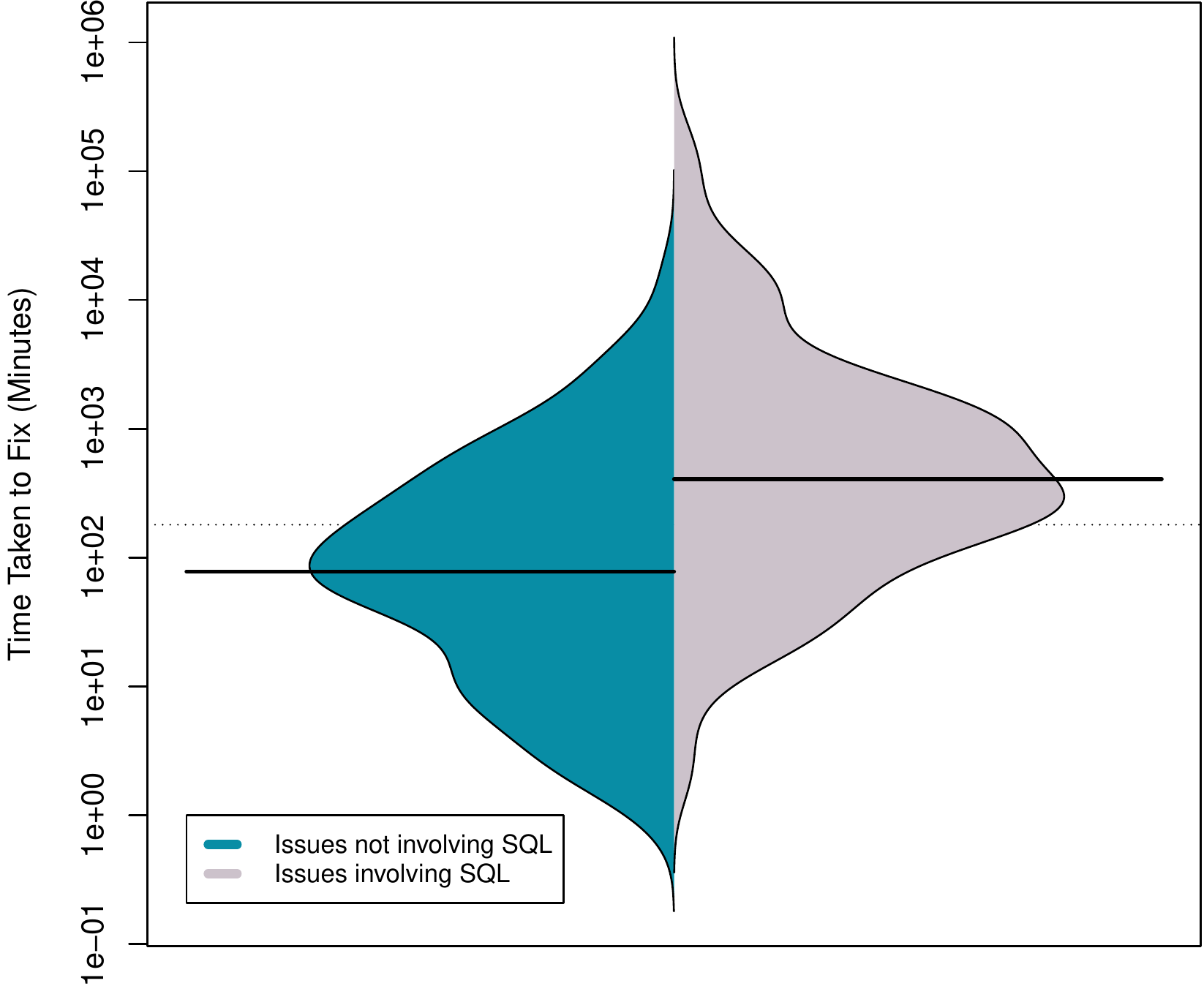}
	\caption{The size of changes (additions plus deletions) of our representative sample \label{fig:sample:changesize}}
\end{figure}

To investigate whether our representative sample maintained the properties of its population (i.e., the 38,160 issue reports), we show in Figures~\ref{fig:sample:timetocompletion} and \ref{fig:sample:changesize} the comparisons regarding time-to-completion and size of changes (which are equivalent to Figures~\ref{fig:timetocompletion:issues} and \ref{fig:churn:issues}). Indeed, the statistical properties hold as both comparisons exhibit a similar behaviour, i.e., $p=2.53x10^{-9}$ with $d=0.27$ (small) for time-to-completion and $p=3.43x10^{-24}$ with $d=0.45$ (medium) for size of changes. Therefore, we proceeded with our document analysis. 

\begin{figure*}[htb]
	\includegraphics[width=\textwidth,keepaspectratio]{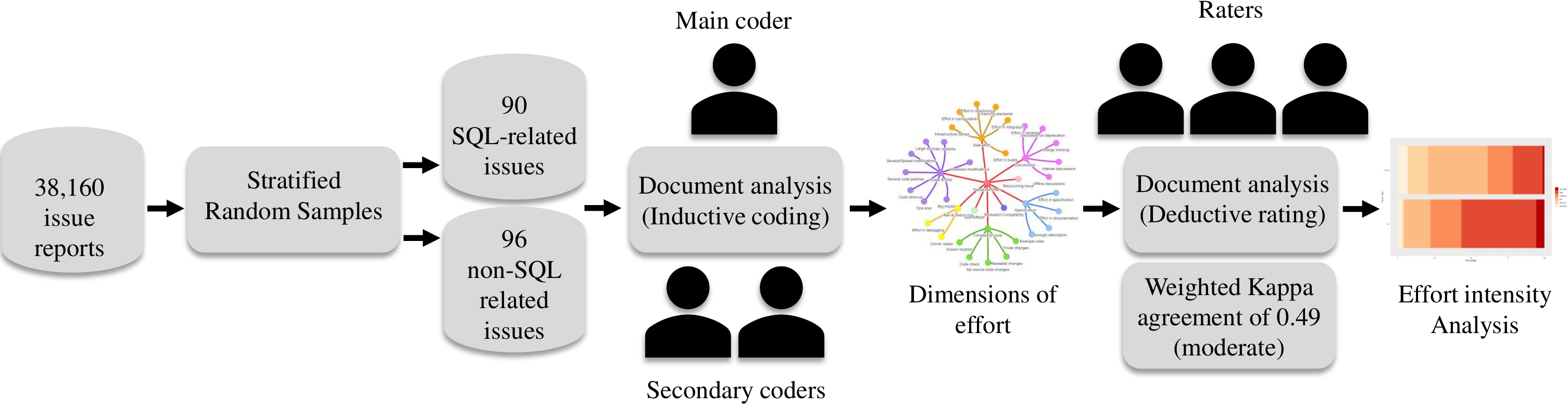}
	\caption{An overview of the document analysis process.\label{rq2:document_analysis}}
\end{figure*}

Figure~\ref{rq2:document_analysis} shows an overview of our document analysis process. Our document analysis consists of two main steps. In the first step, the goal is to find different dimensions of effort invested to address the issue reports within each sample. This inductive approach generates a set of themes related to the different dimensions of effort invested in addressing issue reports. In this step, the first author (``main coder'' in Figure~\ref{rq2:document_analysis}) analyzed all the issue reports from both samples and created {\em themes} based on the observed dimensions of effort invested to address the issue reports. For instance, if an issue report required intense discussions before a solution was proposed (which can be observed in the comments section of an issue report), the theme {\em ``intense discussions''} was created (more details are provided in the results). Once the entire set of themes was generated and documented through several iterations and reflections (see Appendix~\ref{appendix:documentanalysis} for the complete set of themes), the main coder discussed the themes and their meanings with two other authors (``secondary coders''). The secondary coders then used the existing set of themes to code the issue reports from both samples. At this step, the secondary coders had the opportunity to suggest new themes. Next, all coders collaboratively discussed the generated themes (e.g., merging themes, or accommodating new themes), producing a final set of themes. 

It is important to note that our goal with the inductive analysis is to widen our understanding regarding the effort invested in addressing the investigated issue reports instead of finding an {\em absolute truth} regarding what would be the most accurate themes produced in our data. For this reason, we position our inductive approach as a {\em reflexive thematic analysis}~\cite{Braun:2020,Braun:2019} and, hence, we are not interested in {\em coding reliability} at this stage (e.g., measuring inter-rater reliability). Instead, we are interested in generating qualitative results that can help us better reflect on the phenomenon of effort invested in the addressed issues.
 
Once the set of themes is created, the second step is to use the generated themes as a guide for our deductive analysis. The goal of our deductive analysis is to obtain a sense of the intensity of effort invested to address the issue reports. In this deductive analysis, three authors (i.e., three coders) assessed all issue reports separately. Each coder used a five-point Likert scale to indicate their perception of the intensity of effort invested in addressing an issue report; one of ``very low'', ``low'', ``medium'', ``high'', or ``very high.'' In contrast to the previous step, at this stage, we are interested in measuring the accuracy of our perception of exerted effort within an issue report. For this reason, we used the {\em weighted kappa}~\cite{Cohen:1968} as our inter-rater reliability measure. We chose the weighted kappa because it is sensitive to the distance between disagreements, e.g., a disagreement between ``very low'' and ``low'' does not have the same weight as a disagreement between ``very low'' and ``high.'' Finally, we compared the required effort for issues involving SQL code to that of issues that do not involve SQL code.\\

\begin{figure*}[htb]
	\centering
	\includegraphics[width=0.9\textwidth,keepaspectratio]
	{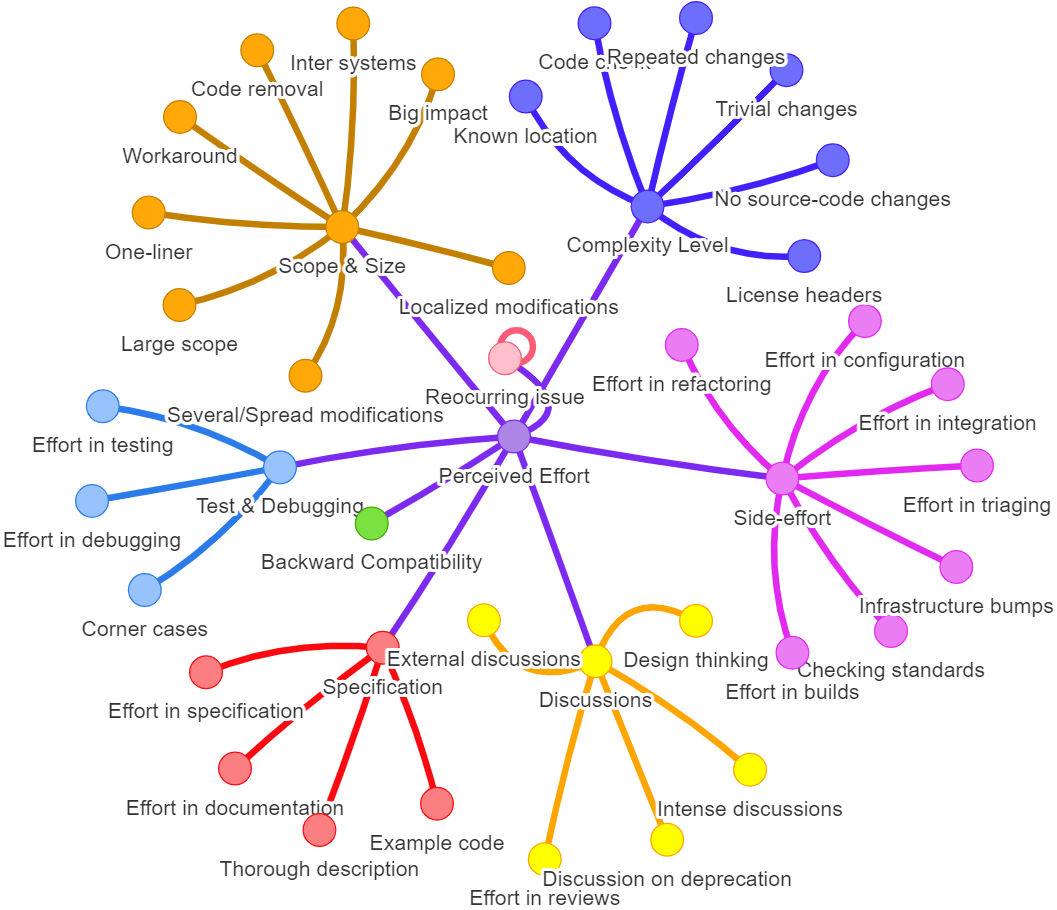}
	\caption{An overview of the dimensions of effort (themes) that emerged from our inductive analysis. \label{fig:dimensionsofeffort}}
\end{figure*}

\begin{figure*}[htb]
	\begin{subfigure}[a]{0.49\textwidth}
	\centering
	\includegraphics[width=\textwidth,keepaspectratio]
	{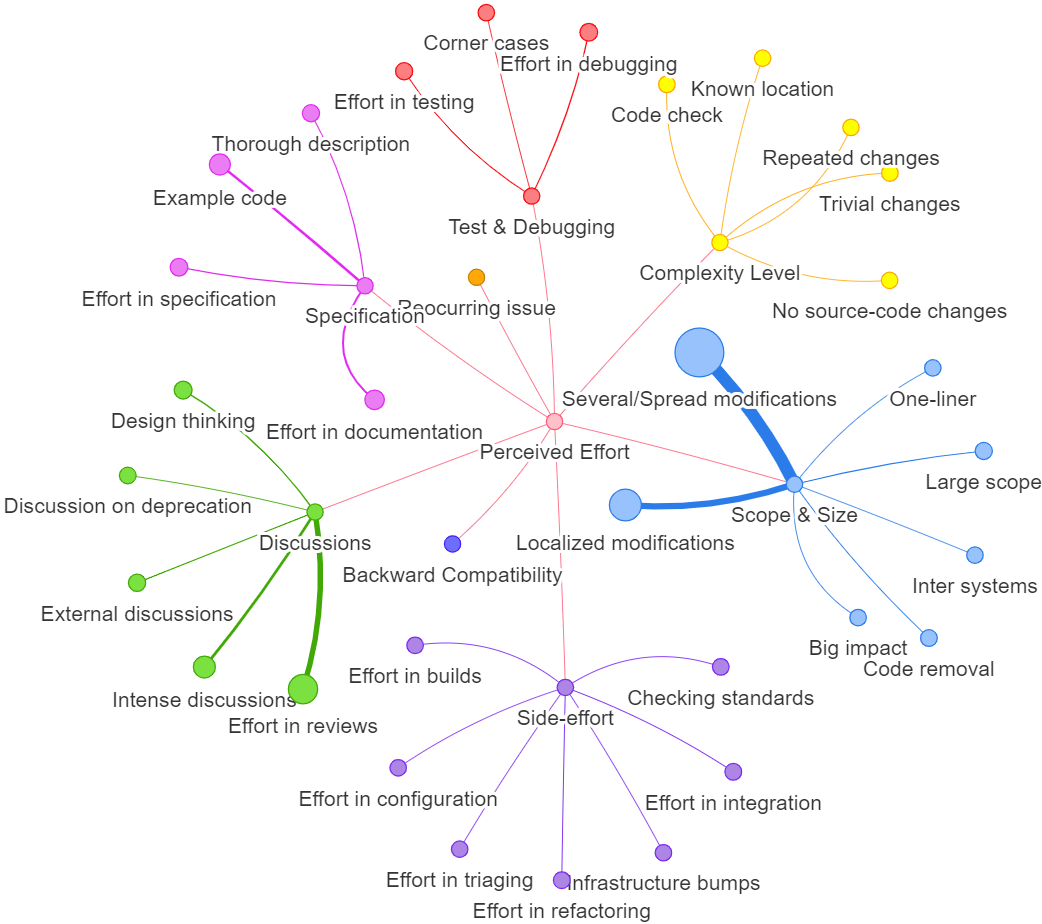}
	\caption{SQL dev-tasks. \label{fig:dimensionspertypesql}}
	\end{subfigure}
	\begin{subfigure}[a]{0.49\textwidth}
	\centering
	\includegraphics[width=\textwidth,keepaspectratio]
	{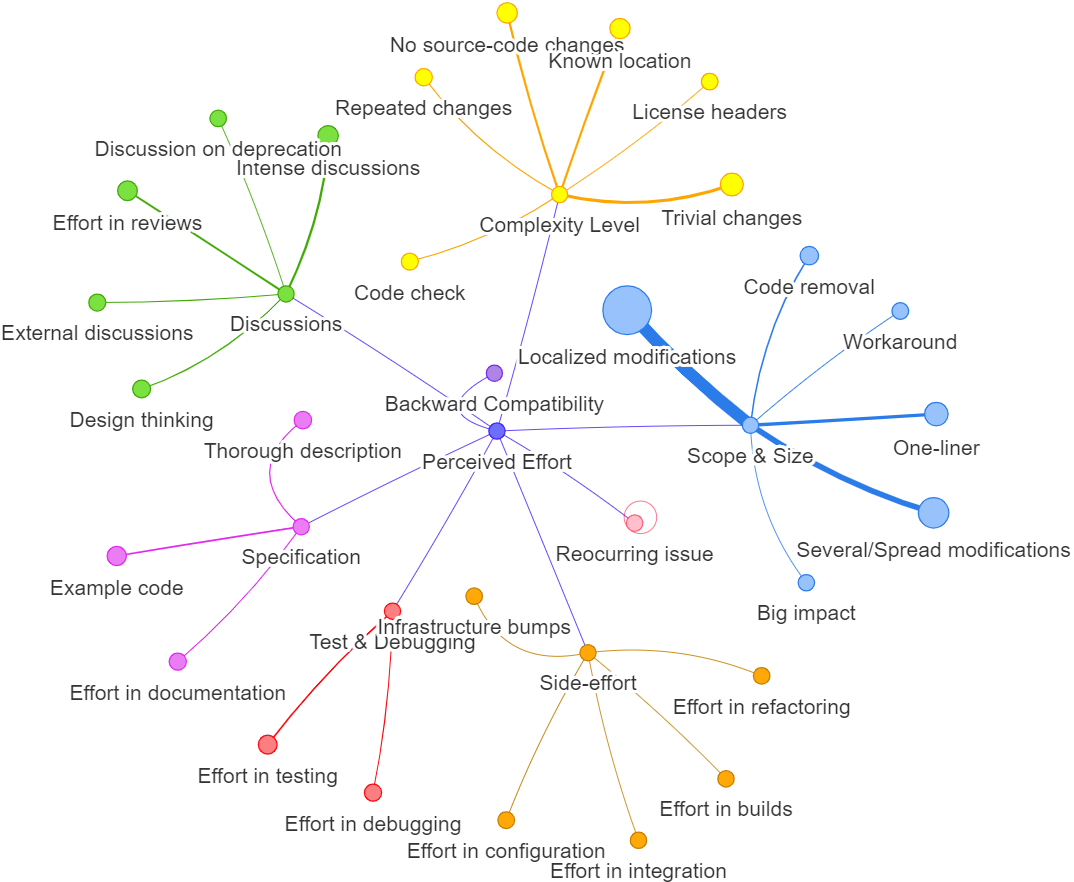}
	\caption{non-SQL dev-tasks. \label{fig:dimensionspertypenonsql}}
	\end{subfigure}
	\caption{Dimensions of effort.}
\end{figure*}

\noindent{\bfseries Results.} Figure~\ref{fig:dimensionsofeffort} visualises the dimensions of effort (i.e., the themes) that emerged from our inductive analysis (see Section~\ref{subsec:rqs}). The central (or root) theme is the \emph{perceived effort}, which is the main object of analysis in this RQ. The second-level (or axial) themes comprise effort related to: \emph{scope \& size}, \emph{discussions}, {\em specification}, {\em test \& debugging}, {\em complexity level}, {\em backward compatibility}, {\em reoccurring issue}, {\em side-effects}, and {\em side-efforts}. The third level themes are more specific and are grouped based on their relationship with the second-level themes. For instance, the theme {\em known location} refers to issues where the developers already knew where to fix the problem from the start. Therefore, the {\em known location} theme falls within the {\em complexity level} axial theme, since already knowing where the problem is located from the start indicates that the issue is not complex after all. Another example is the {\em discussions} axial theme, which groups all themes related to effort invested into discussions (e.g., discussing the design of a solution thoroughly as in the {\em design thinking} theme). In Appendix~\ref{appendix:documentanalysis}, we provide in-depth details about each theme that emerged from our inductive analysis.\\

\noindent\textit{\bfseries SQL development tasks require more widely spread modifications with a larger scope than non-SQL development tasks.}
In the next step of our analysis, we compare the themes for issues that involved SQL code versus issues that did not involve SQL code. Figure~\ref{fig:dimensionspertypesql} visualises the key themes for SQL development tasks. The thickness of the edges and size of nodes are based on the number of times a theme occurred over the total number of theme occurrences.\footnote{\url{https://github.com/neo4j-contrib/neovis.js/}} For example, the theme {\em several/spread modifications} has the highest number of occurrences (188 times) within issues involving SQL code. Hence, the node and edge for {\em several/spread modifications} is the thickest in Figure~\ref{fig:dimensionspertypesql}. Regarding the high-level themes (i.e., the root theme and the Complexity Level, Discussions, Specification, Scope \& Size, Side Effort, and Test \& Debug themes), we consider their number of occurrences as 1 (one), since their role is mostly to communicate how the lower level themes are related to each other.

Figure~\ref{fig:dimensionspertypenonsql} visualises the key themes for non-SQL development tasks. The most notable difference compared to Figure~\ref{fig:dimensionspertypesql} is in the {\em scope \& size} theme: while SQL development tasks require more {\em several/spread modifications} (i.e., the changes involved different files, classes, or packages, or several code changes), non-SQL development tasks require more {\em localized modifications} (i.e., the changes were mostly within a file, or, even if they were in different files, they were only a few), simple {\em workarounds} (i.e., quickly crafted but not ideal solutions) or even {\em one-liner} modifications~\cite{Karampatsis:2020} (i.e., modifications that required only one line of code). This last theme ({\em one-liner}) emerges mostly for non-SQL development tasks. Indeed, because the themes {\em several/spread modifications} and {\em localized modifications} are, in essence, mutually exclusive, we perform a $\chi^{2}$ test of independence~\cite{Chisquare:2013}. Our goal was to check whether there exists a significant difference in occurrences of {\em several/spread modifications} and {\em  localized modifications} between SQL and non-SQL development tasks. We obtain a $p = 1.161\times10^{-18}$ indicating that the observed difference is indeed significant --- Table~\ref{tbl:chisquare} shows our $2x2$ matrix used to compute our $\chi^{2}$ test. Other interesting differences are related to the {\em complexity level} and {\em side-effort} themes. For example, SQL development tasks require different kinds of {\em side-effort} from non-SQL development tasks, such as {\em checking SQL standards}, {\em effort in configuration}, and {\em infrastructure bumps}. For instance, in issue HIVE-15982,\footnote{\url{https://issues.apache.org/jira/browse/HIVE-15982}} a developer comments {\em ``I tested it on Postgres and it agrees with Oracle. So, its [\emph{sic}] worth rechecking the standard for this.''} They then proceed to check the standard and report back on the results. As for the differences in {\em complexity level}, we observe that non-SQL development tasks can involve problems for which the {\em location is known} from the start as well as problems that require {\em no source-code changes}. \\ 

\begin{table}[htb]
	\centering
	\begin{tabular}{lll}
		\toprule
		& {\bfseries Several/Spread modifications} & {\bfseries Localized modifications} \\
		\midrule
		{\bfseries Non-sql} & 89 & 204\\
		{\bfseries Sql} & 188 & 90 \\
		\bottomrule
	\end{tabular}
	\caption{$2x2$ matrix used to compute our $\chi^{2}$ test of independence.\label{tbl:chisquare}}
\end{table}

\begin{figure}[htbp]
	\centering
	\includegraphics[width=.6\textwidth,keepaspectratio]
	{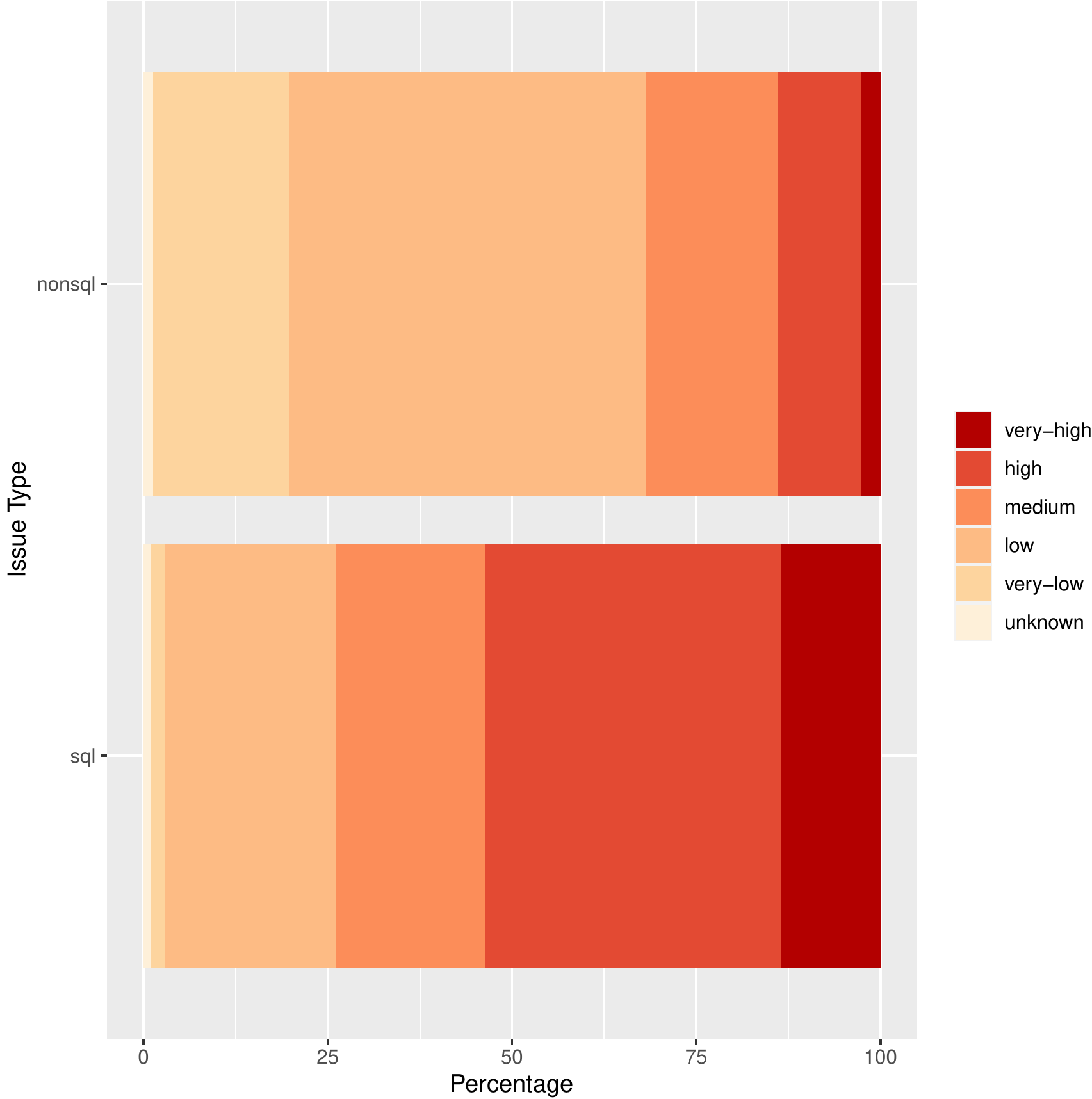}
	\caption{The perceived effort invested in SQL development tasks vs.\ non-SQL development tasks. \label{fig:effortpertasktype}}
\end{figure}

\noindent\textit{\bfseries Development tasks that involve SQL require a higher amount of effort compared development tasks that do not involve SQL.} Figure~\ref{fig:effortpertasktype} shows the result of our deductive analysis through stacked bar charts. We observe that SQL development tasks involve a substantially higher proportion of issues (53\%) rated as ``high'' or ``very-high'' in terms of perceived effort compared to non-SQL development tasks (14\%). In terms of inter-coder agreement, we obtained a weighted Kappa of 0.49, which signifies a good agreement beyond chance \cite{Cohen:1968}.\\ 

\section{Discussion and Implications}\label{sec:discussion}

\noindent{\bfseries Interpreting time-to-completion and change size.} In RQ1, we observed that issue reports involving SQL code have a slightly longer {\em time-to-completion} and a significantly higher {\em change size} when compared to issue reports not involving SQL. In RQ2, we did not observe results that are as conclusive as the results of RQ1, since some projects revealed opposite tendencies in terms of time-to-completion of pull requests. Notwithstanding these observations, one has to be careful when interpreting {\em time-to-completion} and {\em change size}.

In terms of a longer {\em time-to-completion}, the interpretations can take opposite directions. A task may deliberately take longer simply because the task has a lower priority instead of being more complex. As for {\em change size}, while a higher {\em change size} may indicate that more effort has been invested, the nature of the changes also plays a role. For example, even if a high number of changes have been made, if they were mostly copies of the same piece of code (i.e., represented by the theme {\em repeated changes} in RQ3), the higher number of changes is not necessarily indicative of higher effort. We can see a practical example of this issue where the HIVE-1928 issue report received a perceived effort rating of ``medium'' in our analysis in RQ2 despite having a high number of code changes, as most of the changes were repetitive (e.g., changing the function \texttt{priv.getPriv()} to \texttt{priv.toString()} in different locations).\footnote{\url{https://issues.apache.org/jira/browse/HIVE-1928}} Otherwise, we probably would have classified HIVE-1928 as ``high'' perceived effort. 

\begin{figure}[htbp]
	\centering
	\includegraphics[width=.6\textwidth,keepaspectratio]
	{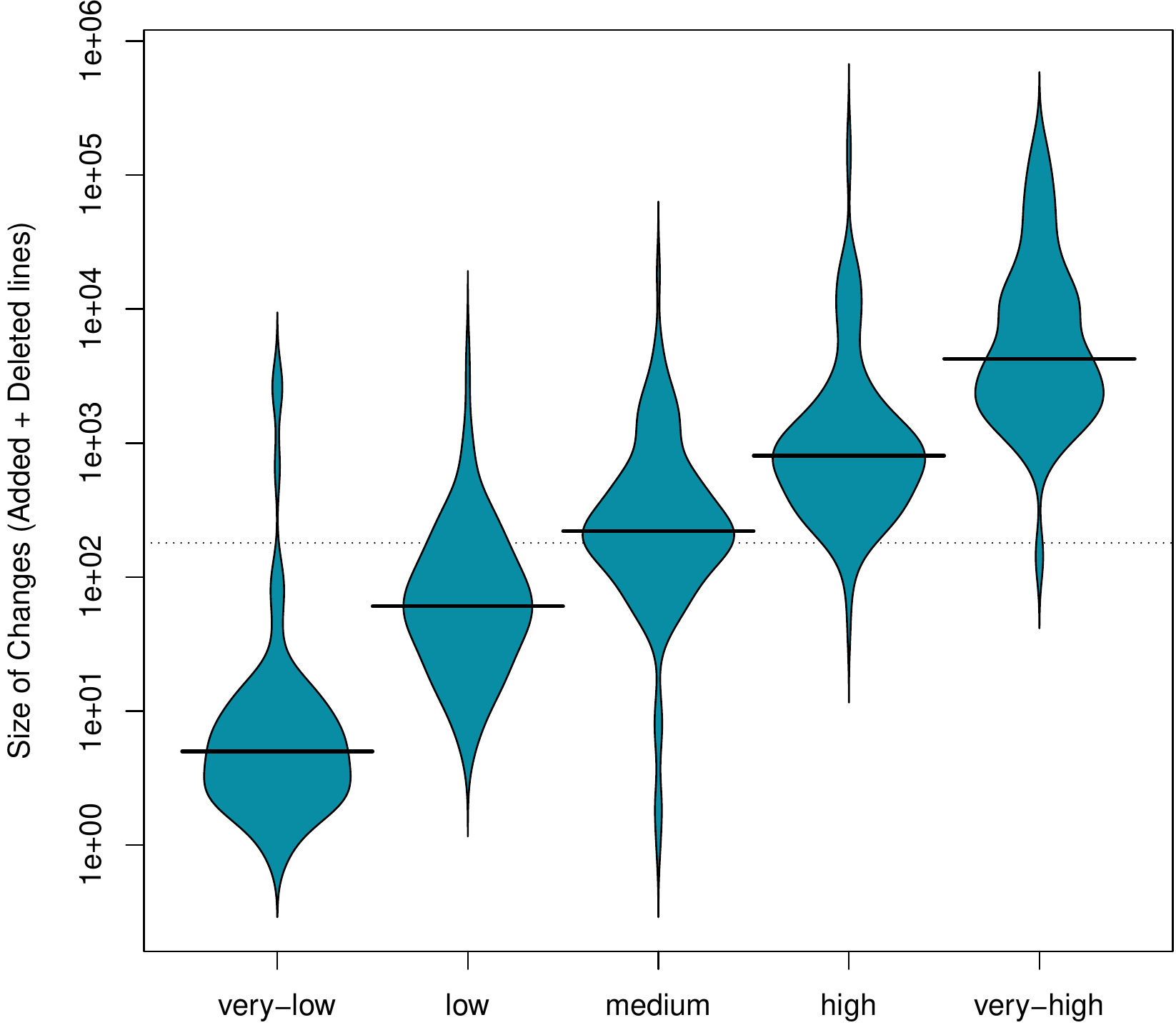}
	\caption{The relationship between the perceived effort and the size of changes of issues. \label{fig:perceivedeffort-vs-churn}}
\end{figure}

Given the prudence required to interpret {\em time-to-completion} and {\em change size}, we further analyse the relationship between our {\em perceived effort} (from RQ3) and the {\em time-to-completion} and {\em change size} of our representative sample. Figure~\ref{fig:perceivedeffort-vs-churn} shows the distributions of change size (y-axis) per category of perceived effort (x-axis).

It is interesting to observe that the size of changes required by issues tend to increase as our perceived effort also increases. Indeed, an issue deemed as requiring a {\em high} effort may need fewer code modifications than a {\em medium} one (as one can note from the variations of the distributions), which can be explained by those issues that, despite not requiring as many code changes, involved more intense discussions or other type of efforts (e.g., effort in reviews). We ran a Kruskal Wallis test~\cite{kruskal1952use} to check whether the different distributions (i.e., very-low, low, medium, high, and very-high) are significantly different from one another. We obtain a statistically significant outcome $p=6.127\times10^{-94}$, indicating that the distributions are likely different. We then ran several pair-wise Wilcoxon tests --- using Bonferroni-Holm corrections to counteract the problem of multiple comparisons) --- and observed that all distributions are statistically different from one another. 

\begin{figure}[htbp]
	\centering
	\includegraphics[width=.6\textwidth,keepaspectratio]
	{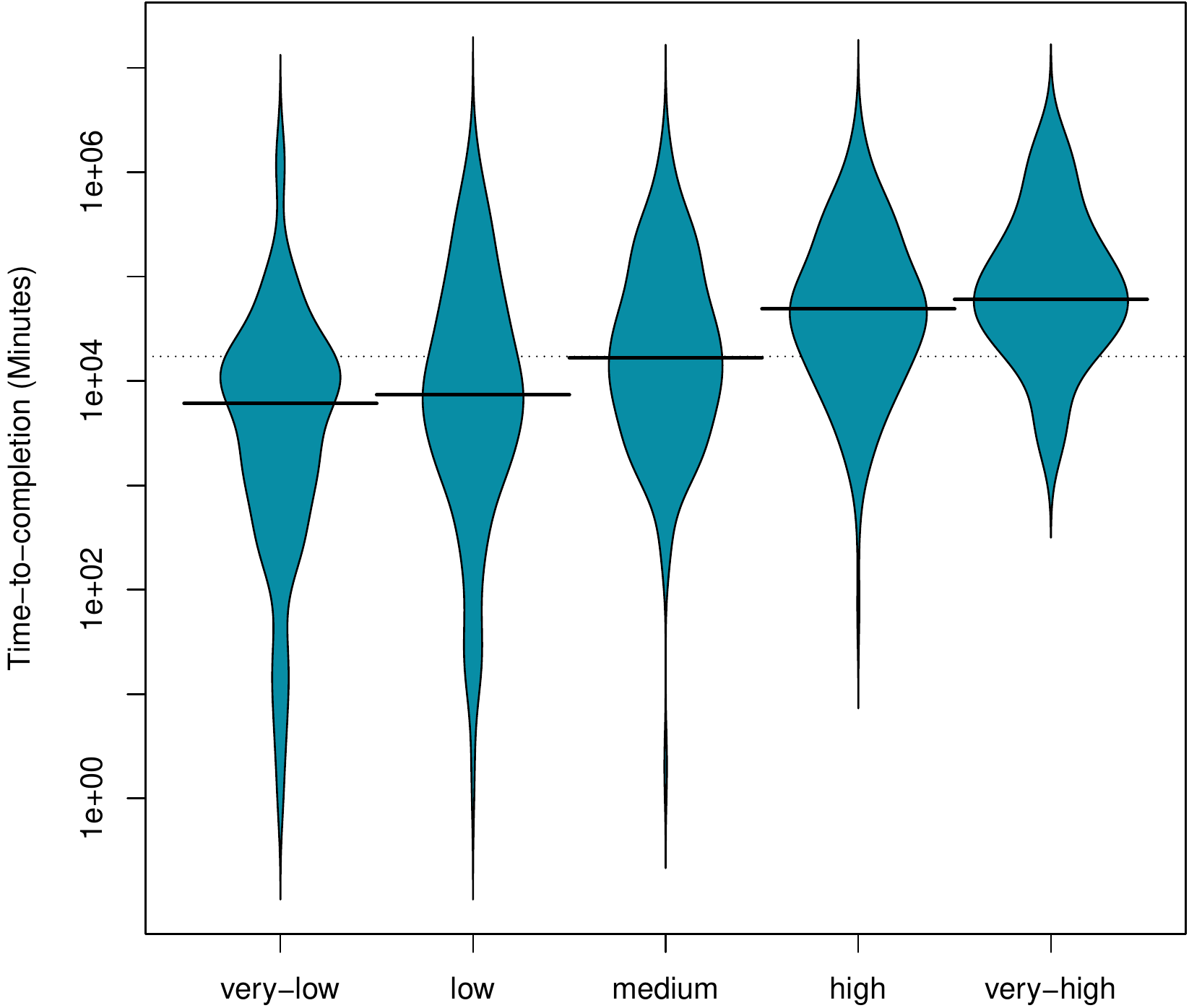}
	\caption{The relationship between the perceived effort and the time-to-completion of issues. \label{fig:perceivedeffort-vs-time}}
\end{figure}

Figure~\ref{fig:perceivedeffort-vs-time} shows the distributions of time-to-completion (y-axis) per category of perceived effort (x-axis). Indeed, the time-to-completion tends to be longer as the perceived effort increases. Again, given the variations in the distributions, a {\em very-low} effort issue could take longer than a {\em high} effort issue, which can be explained by those cases where, although an issue would be easy to fix, the priority of such an issue may not be high. Another interesting observation is that the variation in distributions tends to reduce as the effort increases. For example, it is more likely that a {\em very-high} effort issue will take a longer time to be addressed, whereas a {\em very-low} effort issue may be addressed very quickly or may equally take a longer time.
Our Kruskal Wallis test hints that the distributions are statistically significantly different from one another ($p=4.595\times10^{-24}$). Next, our pair-wise Wilcoxon tests reveal that, when performing pair-wise comparisons, only the {\em very-low vs. low} and {\em high vs. very-high} distributions are not statistically different.

Overall, these result helps us triangulate the results obtained in RQ1. For example, it is less likely that the difference in {\em change size} or {\em time-to-completion} in issue reports involving SQL code is solely due to issues being less important or of a lower priority. Overall, we believe that we have sufficient evidence to conclude that SQL development tasks require more effort from developers.\\

\begin{figure}
	\centering
	\begin{subfigure}[b]{0.49\textwidth}
		\includegraphics[width=\textwidth,keepaspectratio]
		{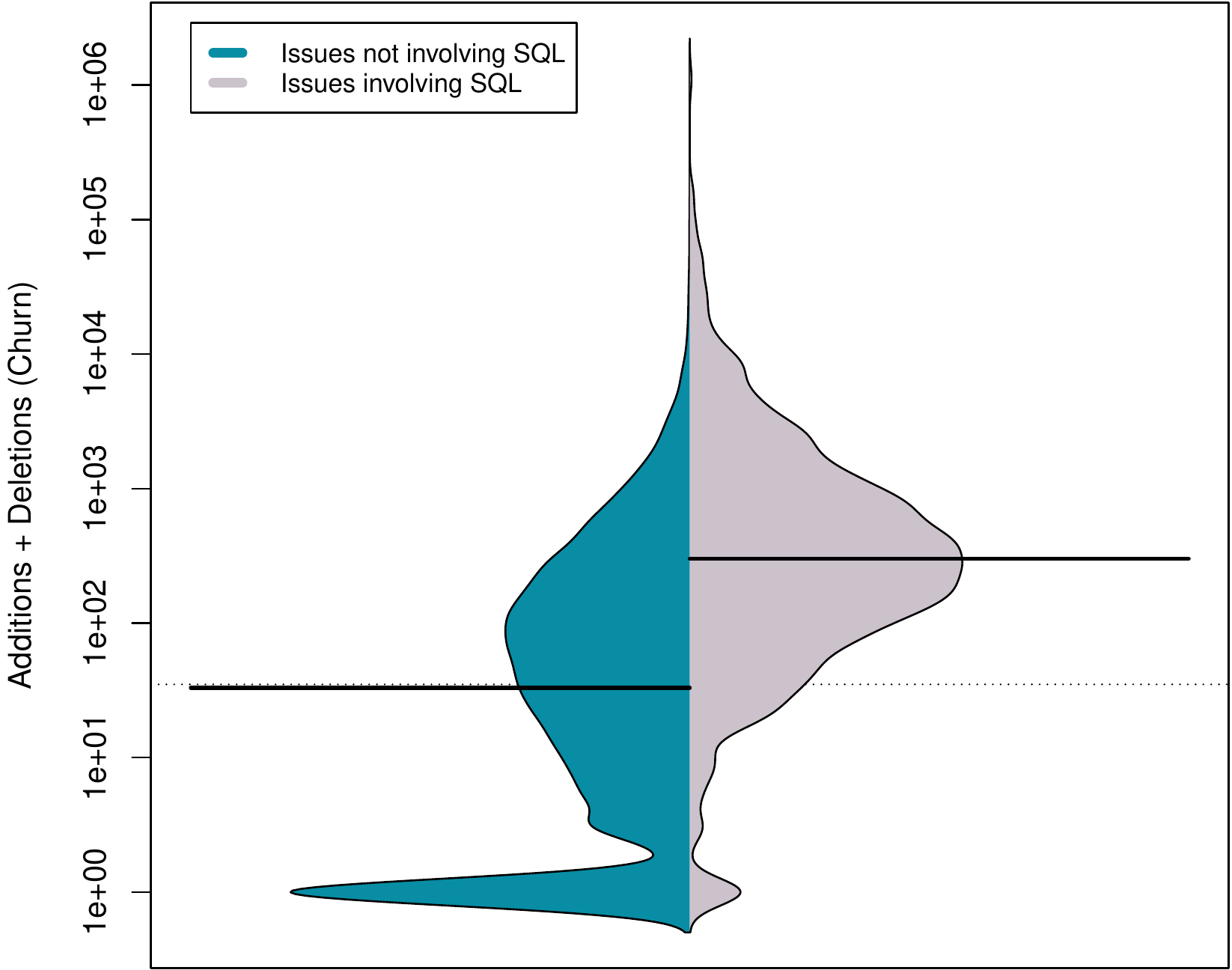}
		\caption{Issues}
	\end{subfigure}
	\begin{subfigure}[b]{0.49\textwidth}
		\includegraphics[width=\textwidth,keepaspectratio]
		{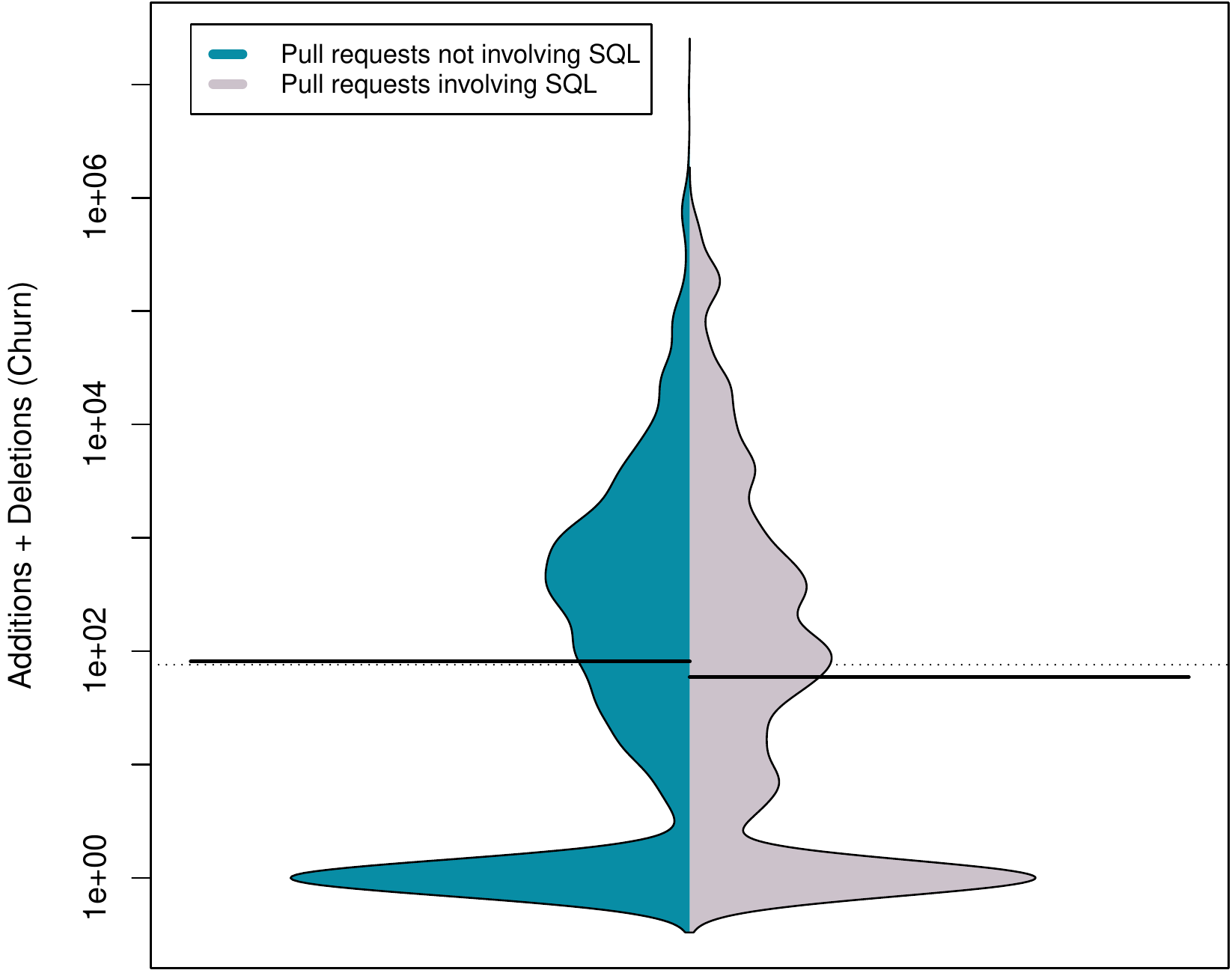}
		\caption{Pull requests}
	\end{subfigure}
	\caption{Distributions of change size for Java-only changes \label{fig:javaonlychanges}}
\end{figure}

\noindent{\bfseries Interpreting the spread-out nature of SQL-related issues.} In RQ3, we observe that SQL-related issues require more changes and/or are more spread-out when compared to SQL-unrelated issues. One might argue that SQL-related issues require more changes not necessarily because of the amount of {\em code} changes involved but because they may co-occur with changes in UI and configuration files (i.e., involving HTML or XML code), which can even involve auto-generated code. Due to this reason, we show in Figure~\ref{fig:javaonlychanges} a comparison of {\em change sizes} similar to RQ1 (issues) and RQ2 (pull requests), but considering Java code changes only.

In terms of issues, we obtain a significant p-value ($p=1.071\times10^{-222}$) with a {\em large} effect-size ($d=0.55$), meaning that the discrepancy in the difference between SQL-related issues and SQL-unrelated issues actually increased when compared to the difference observed in RQ1 (see Figure~\ref{fig:timetocompletion:issues}). Regarding pull requests, we obtain an insignificant p-value ($p=0.789$), which is a comparable result to the one observed in RQ2.\\

\noindent{\textbf{Implications for practice and research.}}
The general trend of our results has implications for practice. For instance, developers, being aware that changes involving SQL code may take longer (or need more coding), may prioritise these changes accordingly (e.g., when prioritising the tasks to tackle in the next sprint). More developers may be added to these tasks, or they may be tagged for enhanced oversight. Another implication of our results is that, knowing that SQL-related tasks may involve more code changes, developers or project managers can consider this information when {\em creating and communicating software estimations} (e.g., when playing planning poker, rules may be carefully followed to ensure the most reliable estimates are entered on the system). The software estimation aspect is particularly important because accurate estimations are directly related to the delivery of a successful software project~\cite{whigham2015baseline,shepperd2014cost}. Additionally, if an issue is likely impacting SQL code, developers may consider this information in their decision to prioritise the code review of the changes involved in such an issue. 

Regarding implications for research, our work opens interesting avenues for future studies and the development of approaches. For example, research has been invested in predicting which issues and pull requests should be addressed next (i.e., prioritisation) ~\cite{zhao2019improving,van2015automatically}. It would be interesting to investigate whether the presence of SQL-code in a patch associated with an issue (or in a pull-request) could help such models to improve their predictions. While this has not been previously considered a pertinent issue, our evidence suggests that it is worth consideration, at least for evaluating relevance. 

Lastly, when considering other contexts, such as in open source development, contributors should be patient that SQL-related issues may take a bit longer or may require more changes to be addressed. Evidence for RQ3 somewhat supports the added challenges presented by SQL-code, including at times the need to consult with SQL standards. This has implications for times when such standards are not readily available too, and so, teams should organise to ensure that project resources are visible to everyone to reduce delays. 

\section{Threats to Validity}\label{sec:threats}

In this section, we discuss the threats to validity of our work.

{\bfseries Construct validity} concerns how our conclusions are based on our evidence. The main construct threat of our work is related to the information that we {\em cannot see}. For example, as discussed in Section~\ref{sec:discussion}, the results obtained in RQ1 can have several interpretations, one of which is that issue reports may take longer to be addressed because they are less important. However, it is challenging to gauge the importance of an issue report. For instance, developers do not always express their opinion about the importance or urgency of an issue report, which makes it harder for researchers to extract this information from the data that is available. For example, in issue GROOVY-3832, a developer commented {\em ``John, any chance you'll be able to look at this in time for 1.7.2 (approx 1 week away) otherwise I'll attempt to take a look.''}\footnote{\url{https://issues.apache.org/jira/browse/GROOVY-3832}} Such a statement can be indicative of some urgency, but it is challenging to identify the degree of urgency or whether the absence of such a statement in other issue reports is indicative of less urgency. To mitigate the limitations related to construct threats, we use both quantitative and qualitative analyses, including inductive and deductive approaches as explained in Section~\ref{subsec:rqs}.

An additional construct threat is the regular expressions we developed to capture SQL statements within source code repositories, which support our investigations throughout this work. These regular expressions may not return all the SQL code in every project (false negatives) and they may return some false positives, which are pieces of code identified as SQL without actually having any SQL code. However, based on the high precision (96.5\%) and recall (97.9\%) of our reliability checks, we expect this risk to be low. 

Another threat is related to the relevancy of the SQL code within commits. For example, one may question whether the SQL code included in the commits was relevant to the changes represented by these commits or whether the inclusion of the SQL code in the patch was inconsequential (e.g., the changes in
the commit spanned areas with SQL code, but did not actually change the logic of the queries). In this regard, we note that 80\% of the SQL-related tasks in our data (i.e., this is the case for both issues and pull requests) contain direct changes to SQL code (i.e., SQL code was directly involved in
deletions/additions/modifications). On the other hand, for the cases where the SQL statements were not directly involved in modifications (i.e., they were present in the patch but within context lines), there is no direct way to determine whether the changes were completely inconsequential, unless
we were able analyse the patches case by case and had the domain knowledge of the studied projects to do so. 

Nevertheless, to account for this threat, we examined 72 of the issues (i.e., a sample with 95\%
confidence level and 10\% confidence interval) where SQL statements were not directly involved but were present in the context lines of the fixing patches. We observed that for 90\% ($\frac{65}{72}$) of these issues, developers are required to understand the SQL statement in the vicinity, meaning that these
changes fixing these issues can still be regarded as SQL-related changes. For example, in change
\texttt{a206ef9ea1126cb4ab3239853633546e93b6f3f8}, which fixed issue \textsc{HIVE-862}, a developer introduced the code \texttt{d.run(cmd).getResponse()}, where \texttt{cmd} is a string holding a \texttt{SELECT} query. It is reasonable to conclude that the developer must understand the
query as they introduced code that executes such a query. Another interesting observation is that 19\% ($\frac{14}{72}$) of these issues have fixing commits that actually directly changed SQL statements but were not identified as such by our regular expressions. For example, in change
\texttt{b6218275b00b64aed7efaf470784cc0441464f67}, which fixed issue \textsc{HIVE-4924}, a developer modified a multi-line SQL-query. However, because the beginning of the query, (i.e., the part containing the \texttt{SELECT * FROM}) was not involved in the change, our regular
expression did not identify such a change as directly modifying an SQL statement, but as having an SQL statement in its vicinity instead. As such, we are confident that our results are unlikely to suffer from bias and are properly based on true SQL-related changes.

Lastly, we acknowledge that addressing pull requests and addressing issue reports can be an interconnected process, i.e., before marking an issue report as {\em addressed} a developer may need to merge a pull request related to that issue report. Nevertheless, for the purpose of our research, we studied issue reports and pull requests separately because our investigated projects do not use pull requests and issue reports in a consistent manner. For example, in the \textsc{Geode} project, developers tend to link a pull request to an issue report.\footnote{See the ``links to'' section on \url{https://issues.apache.org/jira/browse/GEODE-4894}} However, in the \textsc{CloudStack} project, commits are performed directly to the code-base and are mentioned in issue reports, i.e., not necessarily via pull requests.\footnote{See comments section of \url{https://issues.apache.org/jira/browse/CLOUDSTACK-4146}} Thus, although there may be an overlap between the tasks represented by issue reports and pull requests, we analysed issue report and pull requests separately to avoid the intricacies of different projects' processes.

{\bfseries External validity} concerns the extent to which we can generalize our results to other environments (e.g., other software projects). A clear limitation of our research is that we only used Apache projects to conduct our study. Because of this, our conclusions are less generalizable to a wider population of projects and communities. For instance, it may be the case that because Apache projects adhere to a certain rigour in their development process, their SQL code is more stable than in projects of other communities. This potential higher stability of SQL code may influence our results, since SQL code would only be touched by tasks of higher impact and scope. 

However, we weighed the choice of studying only Apache projects against the issues of selecting \textsc{GitHub} projects more randomly and the possibility of selecting small or unusual projects. An interesting future work would be to investigate a greater variety of projects or to compare and contrast two different OSS communities. Different open-source communities may have different practices. For example, in the Mozilla open-source community, they specifically mark performance bugs, whereas in the Apache community there is no option for this in JIRA \cite{Jin:2012}.

Another limitation is focusing only on OSS communities. It is possible that closed software communities have entirely different practices altogether. For example, closed source communities tend to have a few dedicated testers \cite{Bachmann:2009} rather than the more crowd-sourced approach to testing that is seen in OSS projects. This could mean issues are not fixed as quickly or pull requests are not merged as quickly, thus producing different results to our research.

Lastly, although our Apache projects have a substantial number of SQL queries, they cover very different domains. For example, SQL-related tasks developed for Groovy (a programming language) may be different in nature from SQL-related tasks developed for Hive (a data warehouse project). Nevertheless, in this study, we are interested in studying the general trends in the characteristics of SQL-related tasks. Studying whether the characteristics of SQL-related tasks are different for projects of different domains is an interesting possibility for future work.

{\bfseries Internal validity} is generally concerned with the potential confounding factors involved in the relationship between independent and dependent variables (especially in the case of causal relationships). 

In this research, we are not necessarily interested in investigating causal relationships between a dependent variable and its explanatory variables. Instead, our study is more exploratory in the sense that our aim is to better understand the characteristics of SQL development tasks (i.e., tasks that involve both SQL code and application code). In RQ1 and RQ2, the dependent variables were {\em change size} and {\em time-to-completion} and the independent variables were {\em SQL development task} or {\em non-SQL development task}. Nevertheless, we refrain from drawing a causal relationship in RQ1 and RQ2, tapping into more qualitative investigations in RQ3 to help us collect enough evidence to explore the relationship between SQL development tasks and development effort.

Regarding the qualitative analyses, especially those employed in the inductive analysis of RQ3, there is always the potential for bias from the authors' subjective experiences. For example, had the authors had different experiences, the themes that emerged in RQ3 might have been different. Despite this potential threats we employed rigorous qualitative methods to conduct our analyses (e.g., recruiting several coders and computing inter-rater agreement measures). Regarding the inductive analysis in RQ3, we acknowledge that the emergent themes may not be exhaustive and should be taken as a base framework that other researchers can further improve or refine.

Lastly, we acknowledge that, as the authors of this work, we do not have the domain knowledge of the investigated Apache projects in order to fully understand the code-base and the decisions made in the design of these projects. Nevertheless, we are able to search for ``clues'' that indicate the effort invested in an issue. Such clues are present in our scheme described in Appendix~\ref{appendix:documentanalysis}. Such a scheme for our manual analysis was developed through our document analysis process described in the approach section of RQ3. To conclude, although our approach is not perfect (in the sense that we do not have the same technical or domain knowledge that a developer from our chosen projects would have), we are still able to gauge the relative required effort when comparing issues.

\section{Conclusion}\label{sec:conclusion}

The goal of our research is to better understand the characteristics of SQL development tasks (i.e., development tasks involving both SQL code and application code). For this purpose, we set out to conduct an empirical study using 20 Apache projects to investigate whether SQL development tasks have a longer {\em time-to-completion} or whether they require more code modifications (as indicated by a larger \emph{change size}) when compared to other development tasks. Indeed, we observe that SQL development tasks may take slightly longer to be completed and likely require more code modifications. To dive deeper into these findings, we employ a document analysis strategy with two main steps: an inductive analysis and a deductive analysis. As a result of our inductive analysis, we observe several dimensions of effort related to SQL development tasks and non-SQL development tasks. The major difference in terms of dimensions of effort is that SQL development tasks require more spread out changes and extra effort related to SQL-specific tasks such as checking SQL standards. According to our deductive analysis, SQL development tasks also require more effort from developers in general. 

Overall, our empirical research suggests that SQL development tasks not only require extra effort, but also different types of effort. As potential implications of our research, we envision that software development practitioners may take into account the nature of SQL development tasks when performing effort estimations (e.g., when using planning poker in agile) or planning code reviews. Furthermore, those awaiting fixes for SQL-related issues should anticipate the involvement of more rigour from reviewers. As future research, we plan to investigate the interplay between developers' experience and the effort required in SQL development tasks. There is also scope for us to replicate this study using data sets from other communities.

\bibliographystyle{spmpsci}      
\bibliography{mybibliography}   

\appendix

\section{Dimensions of Effort: Themes}\label{appendix:documentanalysis}

	Below we explain the meaning of each of the themes found in the inductive analysis of our {\em document analysis} (as explained in Section~\ref{subsec:rqs}). The themes outlined below along with their descriptions served as a guide to perform our deductive analysis, i.e., gauging the required effort that was necessary to address a development task. 

\begin{description}
	\item	[\bfseries Several/spread modifications.] Whether the changes involved different files (e.g., within different classes or packages) or whether there were many code changes.  
	\item	[\bfseries Several code patches.] Sometimes fixes involve several patch iterations. This might indicate a higher effort invested in the fix. However, several patches with trivial changes do not indicate a high effort. 
	\item	[\bfseries Code check.] Whether the fix involved adding only a code check.
	\item	[\bfseries Debug effort.] Whether there was a considerable effort related to debugging the problem. This criterion can be mostly verified in the comments.
	\item	[\bfseries Intense discussions.] Sometimes the issue does not require large code changes, but the discussions revolving around the changes were intense (i.e., discussing alternative solutions or ideas). This may indicate that the solution required a high effort (even if not many lines of code had been changed).
	\item	[\bfseries Localized modifications.] The opposite of {\em Several/spread modifications}. This criterion indicates that the changes were mostly local (within a file), or, even if they were in different files, they were only a few. This criterion normally denotes ``low'' effort.
	\item	[\bfseries Offline discussions.] Sometimes there is evidence that further discussions were taken offline and not registered in the issue report itself. This criterion may help in judging whether extra effort was required to fix the issue.
	\item	[\bfseries Effort in documentation.] Sometimes, addressing certain issues requires not only new source code, but also new documentation. This criterion represents the effort to produce new documentation. 
	\item	[\bfseries Effort in reviews.] This criterion denotes when higher than usual effort was invested in code review (e.g., they explicitly mention an external system where reviews were held and many comments were placed there). 
	\item	[\bfseries Backward compatibility.] This criterion denotes when extra effort was needed to ensure backward compatibility.
	\item	[\bfseries Specification effort.] This criterion denotes when extra effort was needed to elaborate on the problem at hand. For example, when a developer does not fully understand the problem at hand, they might ask for more clarification. Consider the following quote as a practical example of this criterion: ``What's VC, by the way? Venture Capital?''\footnote{\url{https://issues.apache.org/jira/browse/HIVE-8186}} The meaning was actually ``virtual columns''.
	\item	[\bfseries Effort in configuration.] This criterion denotes when there was more than usual effort in creating or changing the configuration of the system.
	\item	[\bfseries Effort in integration.] When there was more than usual effort required to integrate things (e.g., the software suddenly broke because of the new patches or conflicting merges). 
	\item	[\bfseries Big impact.] This criterion emerges whenever there was an explicit indication that the feature or bug fix would have a large impact on the current system. For example, the bug fix or new feature would require a carefully thought out change.
	\item	[\bfseries Checking standards.] Whenever there is effort in checking standards in order to address the issue. For example, if the issue requires a new SQL function to be developed, developers may need to check what the SQL standard says prior to deciding on a solution.
	\item	[\bfseries Corner cases.] Whenever there was effort in covering corner/specific cases (usually in tests) to increase the quality of the fix or the new feature.
	\item	[\bfseries Infrastructure bumps.] Whenever there was a problem in the infrastructure (not necessarily physical infrastructure, but also soft infrastructure, such as server containers, or database servers) because of the issue. This criterion denotes that there was more than usual effort in fixing/updating the infrastructure.
	\item	[\bfseries Trivial changes.] This criterion is not much related to the size of the change, but denotes when a change was ``easy'' to apply. For example, to address an issue, the developers exposed information that was already computed but had not been exposed before. Another example is the addition of getters/setters, i.e., although the code changes can be large, adding getters/setters is trivial and often automated by IDEs. This criterion normally denotes ``low'' effort.
	\item	[\bfseries Effort in refactoring.] Whenever there was extra effort in performing refactoring operations while fixing/addressing the issue.
	\item	[\bfseries Large scope.] Whenever an issue required changes to several modules of the system. This criterion is an exacerbated version of {\em several/spread modifications}.
	\item	[\bfseries Design thinking.] Whenever there are clear discussions related to the design of a fix or new feature. For example, one developer might argue that an inner class should be sufficient to fix something, but another developer argues that actually creating a new package would be best. 
	\item	[\bfseries Unknown.] Whenever there is not enough information on the issue report (normally this criterion is only used when there are really no comments within an issue report).
	\item	[\bfseries Inter systems.] Whenever the issue require developers to think of the communication/interaction/integration of diverse systems in order to propose a solution.
	\item	[\bfseries Thorough description.] This criterion denotes when there was more than usual effort to describe the issue at hand. For instance, a long description was provided with code examples, etc.
	\item	[\bfseries Discussion on deprecation.] Whenever there is additional effort in discussing whether certain functionalities should be deprecated or not. 
	\item	[\bfseries Effort in builds.] Whenever there is additional (and more than usual) effort in creating/modifying the build system (or running the builds) of the software project.
	\item	[\bfseries One-liner.] This criterion denotes when the issue was addressed with only one line of code. This is an exacerbated version of {\em localized changes}. Other variations of this criterion are {\em two-liners} or \emph{three-liners}. This criterion normally denotes ``very-low'' effort.
	\item	[\bfseries Side effects.] This criterion denotes when there was an additional problem/bug because of the solution that was being developed for the issue at hand. Consider the following excerpt as a concrete example: ``I found a hidden bug in FieldSortedHitQueue that materialized when writing a TestCase for a SortField.BYTE sorting with custom parser (it took me a long time to find out whats [\emph{sic}] happening).''\footnote{\url{https://issues.apache.org/jira/browse/LUCENE-1478}}
	\item	[\bfseries Example code.] This criterion applies whenever there was extra effort in providing codes examples in a discussion of an issue report.
	\item	[\bfseries Repeated changes.] Although code changes may be numerous, they may only be repetitions of the same code, which does not require as much effort. This criterion normally represents lower effort.
	\item	[\bfseries Code removal.] This criterion normally goes along with the {\em known location} criterion. {\em Code removal} normally describes when an issue was addressed by only removing code---normally the developer already knew the location of the code to be removed. This criterion denotes ``low'' effort. 
	\item	[\bfseries No source-code changes.] This code refer to issues that were addressed without touching any source code. This criterion normally denotes ``low'' effort.
	\item	[\bfseries Known location.] When the developers already knew exactly where to change the code to fix an issue. This criterion normally denotes ``low'' effort.
	\item	[\bfseries Reoccurring issue.] When there is evidence that an issue keeps reoccurring even after effort invested in potential fixes.
\end{description}

%
%

\end{document}